\documentclass[dvips]{acta}
\usepackage{supertabular,lscape,epsfig}
\usepackage{amssymb}
\usepackage{amsmath}
\DeclareSymbolFont{ppa}{OT1}{ppl}{m}{it}
\DeclareMathSymbol{\vv}{\mathalpha}{ppa}{'166}

\newfont{\hb}{rphvb at 10pt}
\newfont{\hbo}{rphvbo at 10pt}
\newfont{\bitt}{rptmbi at 12pt}
\newfont{\bits}{rptmbi at 11pt}

\SetPages{33}{65}

\SetVol{62}{2012}

\begin{document}

\newcommand{\TabApp}[2]{\begin{center}\parbox[t]{#1}{\centerline{
  {\bf Appendix}}
  \vskip2mm
  \centerline{\small {\spaceskip 2pt plus 1pt minus 1pt T a b l e}
  \refstepcounter{table}\thetable}
  \vskip2mm
  \centerline{\footnotesize #2}}
  \vskip3mm
\end{center}}

\newcommand{\TabCapp}[2]{\begin{center}\parbox[t]{#1}{\centerline{
  \small {\spaceskip 2pt plus 1pt minus 1pt T a b l e}
  \refstepcounter{table}\thetable}
  \vskip2mm
  \centerline{\footnotesize #2}}
  \vskip3mm
\end{center}}

\newcommand{\TTabCap}[3]{\begin{center}\parbox[t]{#1}{\centerline{
  \small {\spaceskip 2pt plus 1pt minus 1pt T a b l e}
  \refstepcounter{table}\thetable}
  \vskip2mm
  \centerline{\footnotesize #2}
  \centerline{\footnotesize #3}}
  \vskip1mm
\end{center}}

\newcommand{\MakeTableApp}[4]{\begin{table}[p]\TabApp{#2}{#3}
  \begin{center} \TableFont \begin{tabular}{#1} #4 
  \end{tabular}\end{center}\end{table}}

\newcommand{\MakeTableSepp}[4]{\begin{table}[p]\TabCapp{#2}{#3}
  \begin{center} \TableFont \begin{tabular}{#1} #4 
  \end{tabular}\end{center}\end{table}}

\newcommand{\MakeTableee}[4]{\begin{table}[htb]\TabCapp{#2}{#3}
  \begin{center} \TableFont \begin{tabular}{#1} #4
  \end{tabular}\end{center}\end{table}}

\newcommand{\MakeTablee}[5]{\begin{table}[htb]\TTabCap{#2}{#3}{#4}
  \begin{center} \TableFont \begin{tabular}{#1} #5 
  \end{tabular}\end{center}\end{table}}

\newfont{\bb}{ptmbi8t at 12pt}
\newfont{\bbb}{cmbxti10}
\newfont{\bbbb}{cmbxti10 at 9pt}
\newcommand{\uprule}{\rule{0pt}{2.5ex}}
\newcommand{\douprule}{\rule[-2ex]{0pt}{4.5ex}}
\newcommand{\dorule}{\rule[-2ex]{0pt}{2ex}}
\begin{Titlepage}
\Title{Detached Red Giant Eclipsing Binary Twins:\\
Rosetta Stones to the Galactic Bulge}
\vspace*{4pt}
\Author{D.\,M.~~N~a~t~a~f,~~~A.~~G~o~u~l~d~~~and~~~M.\,H.~~P~i~n~s~o~n~n~e~a~u~l~t}
{Department of Astronomy, Ohio State University, 
140~W. 18th~Ave., Columbus, OH~43210, USA\\
e-mail: (nataf,gould,pinsono)@astronomy.ohio-state.edu}
\vspace*{4pt}
\Received{March 24, 2012}
\end{Titlepage}

\vspace*{4pt}
\Abstract{We identify 34 highly-probable detached, red giant eclipsing
binary pairs among 315 candidates in Devor's catalog of $\approx10\,000$
OGLE-II eclipsing binaries. We estimate that there should be at least 200
such systems in OGLE-III. We show that spectroscopic measurements of the
metallicities and radial-velocity-derived masses of these systems would
independently constrain both the age-metallicity and helium-metallicity
relations of the Galactic bulge, potentially breaking the age-helium
degeneracy that currently limits our ability to characterize the bulge
stellar population. Mass and metallicity measurements alone would be
sufficient to immediately validate or falsify recent claims about the age
and helium abundance of the bulge. A spectroscopic survey of these systems
would constrain models of Milky Way assembly, as well as provide
significant auxiliary science on research questions such as mass loss on
the red giant branch. We discuss the theoretical uncertainties in stellar
evolution models that would need to be accounted for to maximize the
scientific yield.}{Galaxy: bulge -- Galaxy: stellar content -- binaries:
eclipsing}

\vspace*{4pt}
\Section{Introduction}
To understand the origin and evolution of the Galactic bulge, one would
like to measure the age, kinematics, and abundances for a large and
representative sample of stars. Because both photometric and spectroscopic
ages are strongly degenerate with helium abundance (Mar{\'{\i}}n-Franch
\etal 2010, Nataf and Gould 2011), it is absolutely essential that each
star in such a sample have an estimate of its helium content, in addition
to the ``metal'' abundances that are more usually reported. This appears
to be a daunting requirement: despite the fact that helium comprises
25--40\% of the baryonic mass of bulge stars, there are no reported
spectroscopic helium estimates, other than that of the stripped B-star S2
orbiting the supermassive black hole at the center of our Galaxy (Martins
\etal 2008). Helium is simply too tightly bound to give rise to detectable
lines in the relatively cool stars that inhabit the bulge, though the
application of Herculean techniques to the HeI 10830 has shown promise in
the case of the metal-poor globular cluster NGC~2808 (Pasquini \etal 2011).

Here we argue that well-detached double-red-giant eclipsing binaries can
provide a large ${\cal O}(10^2)$ sample of such well-characterized
stars. The choice of double-red-giant eclipsing binaries is far from
obvious. Red giants (RGs) of luminosity comparable to the red clump (RC)
are themselves relatively rare, roughly one per $10^3$~\MS. And it is
straightforward to show that only ${\cal O}(10^{-5})$ of these have
detached eclipsing secondaries of comparable size. So, for example, one
does not expect even one such binary in the entire system of Milky Way
globular clusters. Detached eclipsing turnoff stars are at least 1000 times
more plentiful. However, these systems do exist and are accessible once one
has access to photometric surveys as large as OGLE: We note the recent
detailed investigation of 3.5~\MS RG twins in a 371.6~day orbit, OGLE
SMC113.3 4007 (Graczyk \etal 2012).

Nevertheless, for the specific problem of tracing the bulge population,
double-RG eclipsing binaries are very much preferred. The principal reason
is simply that they are brighter and hence all measurements are much more
precise given instrumentation available currently or in the foreseeable
future. For example, currently it is possible to obtain detailed
spectroscopic abundances for turnoff stars only when they are highly
magnified (Bensby \etal 2010, 2011). For the same reason, precision radial
velocity (RV) measurements, needed for accurate masses, are extremely
costly for turnoff stars. Even photometric light curves, which are required
to measure the mean density of the system, are challenging (\eg Clarkson
\etal 2011). This is not just because of the low flux levels of the source
but more fundamentally, irreducible blending by ambient stars, as well as
third bodies that are frequently present in close-binary systems (Tokovinin
\etal 2006, Pribulla and Rucinski 2006). Finally, because the bulge has
finite depth ($\sigma_{(m-M)_{0}}\approx0.15$~mag) and is subject to
differential extinction, one cannot, in strong contrast to globular
clusters, use photometric information to precisely determine the phase of
stellar evolution of a given turnoff star.

All of these problems are greatly reduced in RGs. Plentiful photons easily
yield high signal-to-noise ratio (S/N) measurements. Blending is
intrinsically less important because the sources are bright, and moreover
it is possible to detect blends down to very low flux levels. High flux
levels and low-blending imply that proper motions are more precisely
measured, and precise light curves enable accurate distance measurements
using standard eclipsing-binary techniques (at least to determine relative
positions within the bulge). Finally, the whole phase of RG evolution is
short, so that the phase of stellar evolution can be determined extremely
precisely. The only real problem is the low frequency of detached
double-red-giant eclipsing binaries. However, in this paper we show that at
least 34 such systems are already present in the eclipsing-binary catalog
constructed by Devor (2005) from OGLE-II data.

The structure of this paper is as follows. In Section~2, we show how to
identify Galactic bulge RG EB twins from an order of magnitude larger list
of candidates. The observed frequency of these systems in OGLE-II implies
that there are of order 200 more waiting to be discovered in OGLE-III
data. In Section~3, we summarize the stellar models used in this work. In
Section~4, we discuss the immediate science that could be achieved with
even preliminary spectroscopic follow-up. We find that mass and metallicity
measurements would, by themselves, be sufficient to validate or falsify
several hypotheses of the bulge stellar population. In Section~5, we
investigate how well the system parameters must be measured and how the
precision of these measurements translate into errors on the age-helium
plane. Further, we show that if current theoretical and observational
uncertainties on the RG temperature scale are reduced in the future, and if
tidal effects could be taken into account by either selecting very detached
systems or effectively modeling them, detached RG eclipsing binaries could
in and of themselves be sufficient to fully characterize the Galactic bulge
age-helium-metallicity relationship.

\Section{At Least 34 Detached Red Giant Eclipsing Binary Pairs in the OGLE-II Eclipsing Binary Catalog}
There are significant challenges in finding RG eclipsing binary
pairs. First, the lifetime of the RG phase is only $\approx1\%$ of the
stellar lifetime, which means that the two masses can differ by no more
than $\approx0.5\%$ for both stars to be in the RG phase
simultaneously. Second, from Newton's generalization of Kepler's third law:
$$\biggl(\frac{a}{10~\RS}\biggl)=0.84\biggl(\frac{P}{\rm{2~days}}\biggl)^{2/3}\biggl(\frac{M_1+M_2}{2~\MS}\biggl)^{1/3},\eqno(1)$$
it follows that RG eclipsing binary pairs will not be detached for the
short orbital periods that are the most easily detected, due to their large
physical size. Meanwhile, at longer periods, the geometrical probability of
eclipse will go down as the inverse of the orbital separation, and even for
fortuitous alignments for which $\sin{i}\approx1$, the S/N will drop
sharply with increased orbital separation as there will be fewer completed
periods to phase the light curve over.

\begin{figure}[htb]
\includegraphics[width=13cm]{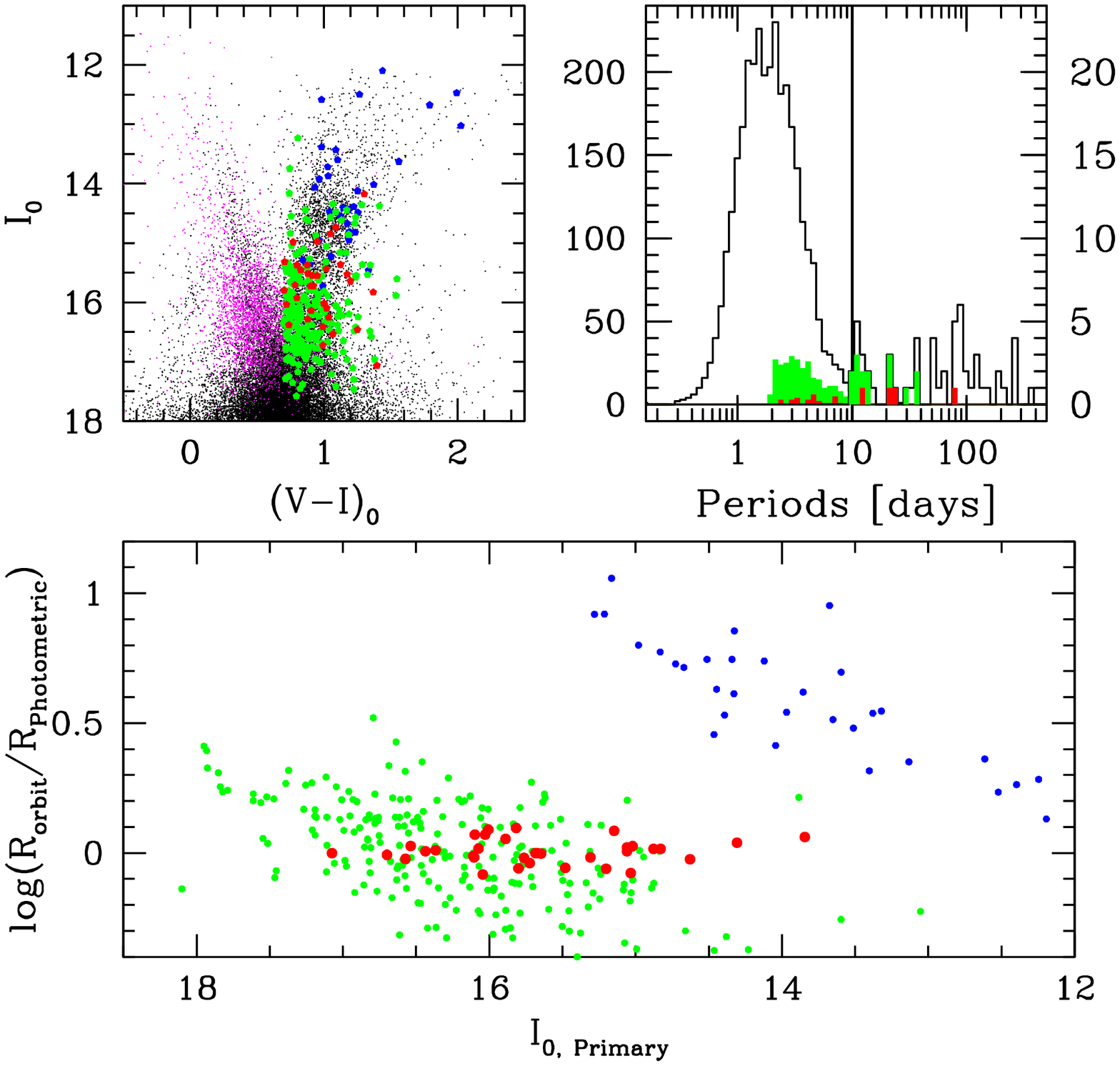}
\vskip4pt
\FigCap{At least 34 of the 10\,862 OGLE-II eclipsing binary candidates in
the catalog of Devor (2005) are detached RG eclipsing binary pairs. {\it
Top-left:} dereddened color and magnitude of detached binaries (magenta),
``gold'', ``silver'', and spurious eclipsing binary sources with
$(V-I)_0\geq0.7$~mag and $P\geq2$~days (red, green, blue respectively),
overplotted on a dereddened OGLE-III CMD toward Baade's window
(black). {\it Top-right:} histogram of period distribution of 3\,170
detached eclipsing binaries, with the distribution of periods greater than
10~days enlarged. {\it Bottom:} comparison of stellar radii derived from
the photometric and orbital information for the primary stars as a function
of $I_{0,\rm{Primary}}$. Same color scheme as {\it top-left}. The gold
candidates pass the test of having consistent radii from photometry and
orbital parameters.}
\end{figure}
It is therefore not surprising that a photometric database as large as that
of OGLE is required to produce a catalog of these systems. Devor (2005)
combed through 218\,699 variable stars in the OGLE-II bulge photometric
survey, to identify 10\,862 eclipsing binaries. There were 3\,170
classified as detached eclipsing binaries, the overwhelming majority of
which are foreground main-sequence disk stars. The top-left panel of Fig.~1
shows the dereddened color and magnitude of the 3\,170 detached eclipsing
binaries overplotted on a dereddened OGLE-III CMD of Baade's window
(Szyma�ski \etal 2011). Only 315 of the catalog members have dereddened
color $(V-I)_0\geq0.7$~mag and $P\geq2$~days.

Though each of these candidates appear as RGs on the CMD and have eclipsing
binary light curves, many are not true RG eclipsing binaries. We find that
at least 33 of these 315 candidates have unphysical parameters, as
determined by deriving radius estimates. The solutions of Devor (2005)
include the orbital periods and the ratios $R_{1,2}/a$, allowing estimates
of the primary's physical radius, $r=5.30\RS(r_1/a)(P/{\rm days})^{2/3}$,
where we have assumed $M_1+M_2=2~\MS$. This assumption of the mass will not
contribute significant error due to the small value of the exponent:
1/3. An independent estimate of the radii of these stars can be derived
photometrically, by de-reddening their colors and magnitudes assuming an
intrinsic color and brightness of the red clump (RC)
$(V-I,I)_{RC}=(1.06,{-}0.10)$, $(m-M)_{0,\rm{bulge}}=14.52$ and using the
empirical color-surface-brightness relations of Kervella \etal (2004), by
first transforming $(V-I)$ to $(V-K)$ using the {\it VIK} color--color
relation of Bessell and Brett (1988). The results of this comparison are
shown in the bottom panel of Fig.~1. There is a sequence of sources,
plotted in blue, for which the ratio
${R_{\rm{orbit}}/R_{\rm{photometric}}}$ reaches values as high as 10.
These are manifestly unphysical, and they comprise $\approx10\%$ of the
eclipsing-binary candidates with $(V-I)_0\geq0.7$~mag and
$P\geq2$~days. That these form a sequence strongly implies that they are a
specific class of variables that happen to strongly resemble eclipsing
binaries. We note that they are all as bright or brighter than the RC.

The 34 red points, which we call the ``gold sample'', have a minimal size
criterion $R_{\rm secondary}\geq3\RS$ (to ensure that both members are RG
stars), and a consistent size criterion
$|\log(R_{\rm primary,orbit}/R_{\rm primary,photometric})|\leq0.10$,
equivalent to a $\approx25\%$ error in the radius or 0.5~mag in the
brightness. This is an estimate of the effects due to errors in the
corrections between color and surface brightness, differential reddening,
and depth relative to the Galactocentric distance.  248 of the candidates,
plotted in green, have properties that are physical but not optimal. Many
are likely disk stars, other may have secondaries on the subgiant branch,
for which follow-up analysis would not be able to assume $M_1=M_2$. We
classify these as the ``silver'' sample. We directly inspected each of the
34 gold eclipsing binary light curves by downloading photometry from the
OGLE-II archive (Udalski \etal 1997, Szyma�ski \etal 2005). The OGLE-II
light curves for the gold-sample candidates are shown in Figs.~2,~3, and
~5. Both ``gold'' and ``silver'' candidates are listed in Tables~1 and~2
(in Appendix).

There may be more such systems to be found at brighter magnitudes. It has
been demonstrated that there is a substantial population of undiscovered,
bright periodic variables toward the Galactic bulge. However, OGLE
photometry, from which the eclipsing binary candidates are derived,
saturates at $I\approx13$~mag (Szyma�ski \etal 2005). Nataf \etal (2009),
as part of a microlensing feasibility study investigating bright
($8\lesssim I\lesssim 13$) stars toward the Galactic bulge, took 151
exposures spanning 88 nights and estimated that 50\% of the periodic
variables were not previously detected. Due to the short baseline, it is
not surprising that no long-period detached eclipsing binaries are present
in the catalog of 52 previously undetected eclipsing binaries (Nataf \etal
2010), but it is likely that a few could be found near the tip of the RG
branch by a dedicated campaign with a small-aperture telescope.

\subsection{Uncertainties in the Photometric Parameters}
We comment on a few uncertainties in the parameters derived by Devor (2005)
and how they can be rectified.

The first is that of ellipsoidal variations, which are not accounted
for. Ellipsoidal variations result from the geometric distortion of
close eclipsing binary stars due to their mutual gravitation. That this
was not taken into account by Devor (2005) likely results in small
errors in some of the parameters such as the stellar densities. However,
this effect could be well approximated for in a more detailed study.

To better understand these effects, we follow the derivation of Assef \etal
(2006). From Eq.~(6) of Morris (1985), the mean magnitude difference
between maxima and minima resulting from ellipsoidal variations,
$\Delta M$, behaves as:
$$\Delta M=0.325\frac{(\tau_1+1)(15+\mu_1)}{(3-\mu_1)}\biggl(\frac{m_2}{m_1}\biggl)\biggl(\frac{R_1}{a}\biggl)^3{\sin}^2i\eqno(2)$$
where $\mu_1$ is the primary's linear limb-darkening coefficient and
$\tau_1$ is the primary's gravity-darkening coefficient. From Table~1 of
Al-Naimiy (1978), we find $\mu_1\approx0.6$, and $\tau_1\approx0.4$ from
Al-Naimiy (1978) and Eq.~(10) of Morris (1985), where we assume
4000~K objects measured at 8000~\AA\ for both parameters. For edge-on
RG EB twins, $(m_2)/(m_1){\sin}^2i\approx1$, and thus Eq.~(2) reduces
to:
$$\Delta M\approx2.96\biggl(\frac{R_1}{a}\biggl)^3=0.024\biggl(\frac{R_1/a}{0.2}\biggl)^3\eqno(3)$$
which corresponds to the typical between-eclipse trends seen in Figs.~3, 4,
and~5.

Some readers may be concerned about a possible factor of 2 degeneracy in
the period calculation: Devor (2005) assumes there are always two
measurable eclipses. We argue that this is a valid assumption for these
stars. First, with their position on the bulge CMD and the match between
their photometric and orbital radii, we are confident that these are RG
stars. Second, the eclipse depths seen in Figs.~3, 4, and 5 range from
0.1~mag to 0.4~mag. It is difficult to conceive of a plausible object that
could eclipse $\approx30\%$ of a RG's light and not be a RG itself, at
which point two eclipses would be inevitable.

The third source of error is that of the eclipse phase. Observers may wish
to know the phase of the orbit to optimize their RV targeting, for example
by obtaining the spectra during the secondary eclipse, when the smaller RG
is fully obscured by the larger RG, an epoch we label
$E_2$. Unfortunately, knowledge of the period phases is now somewhat
lost, since OGLE-II observations (Udalski \etal 1997, Szyma�ski
\etal 2005) were taken in the period 1997--2000, and typical periods for
these sources is 20~days. Additionally, many of our values of $E_2$ may be
off by $\approx P/2$ if the photometric fit incorrectly determined the
surface brightness ratio of the two stars. This phase drift will be easy to
account for once time-series photometry from the OGLE-III survey (Szyma�ski
\etal 2011) become available, as these will allow tighter period
determinations and cover the time baseline 2002--2009.

\subsection{Biases in the Sample of Devor (2005)} 
\begin{figure}[htb]
\centerline{\includegraphics[width=10cm]{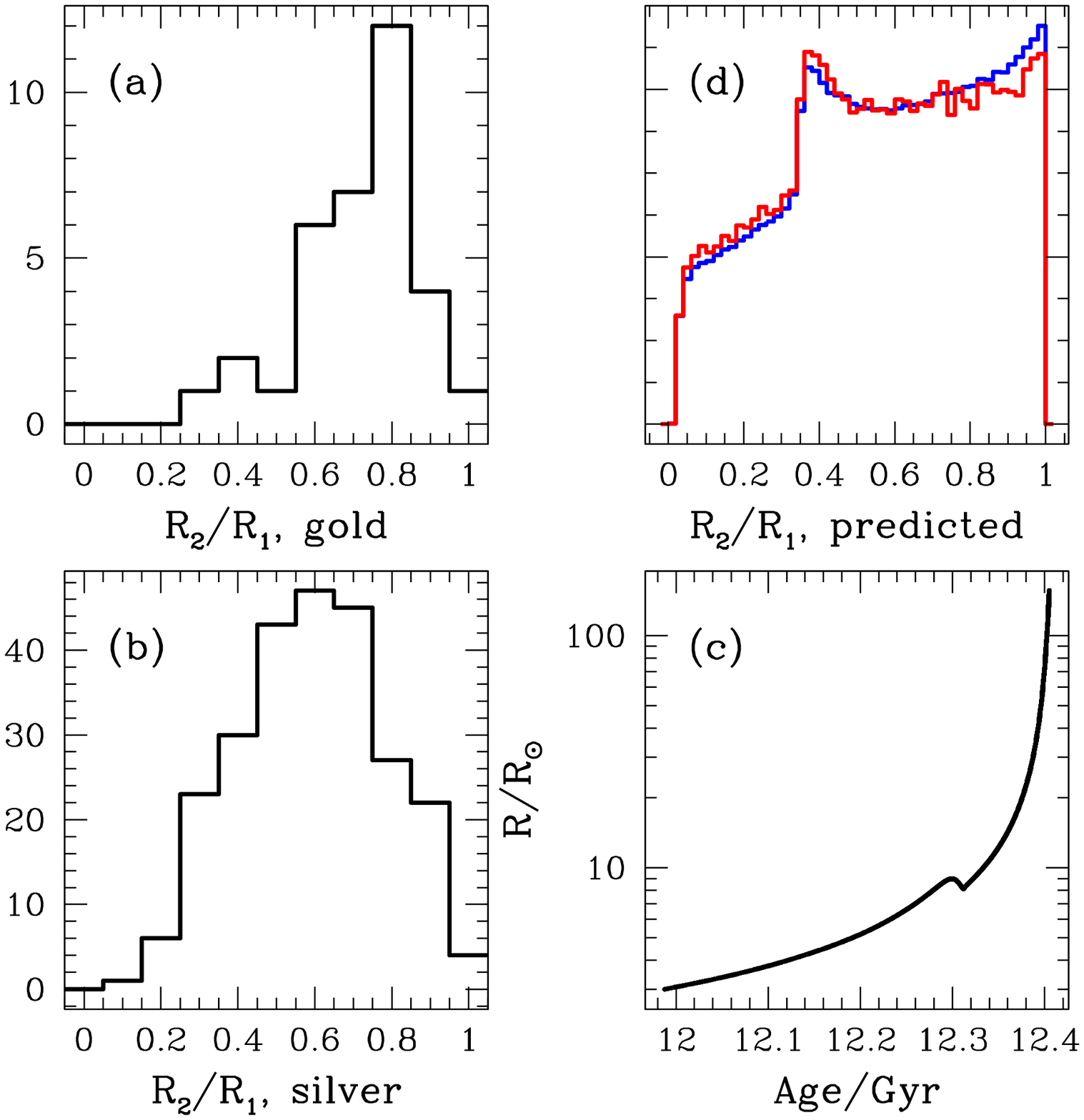}}
\vspace*{4pt}
\FigCap{Observed distribution of $R_2/R_1$ does not match
simple expectations, suggesting selection effects. {\it Panel (a):}
$R_2/R_1$ for the 34 gold sample EB systems. {\it Panel (b):} $R_2/R_1$ for
the 248 silver sample EB systems. {\it Panel (c):} a solar chemistry, 1~\MS
stellar track computed using the Yale Rotating Stellar Evolution Code
(YREC, van Saders and Pinsonneault 2012), we show the track for
$R\geq3~\RS$, up to the tip of the RG branch. {\it Panel (d):} predicted
$R_2/R_1$ for any binary RG twin sampling the probability density function
from {\it Panel c} (blue), and predicted $R_2/R_1$ for detached binary RG
twins modified to account for the period distribution of Duquennoy and
Mayor (1991), and assuming an eclipse probability $(R_1+R_2)/a$ (red).}
\end{figure}
\begin{figure}[htb]
\centerline{\includegraphics[width=13cm]{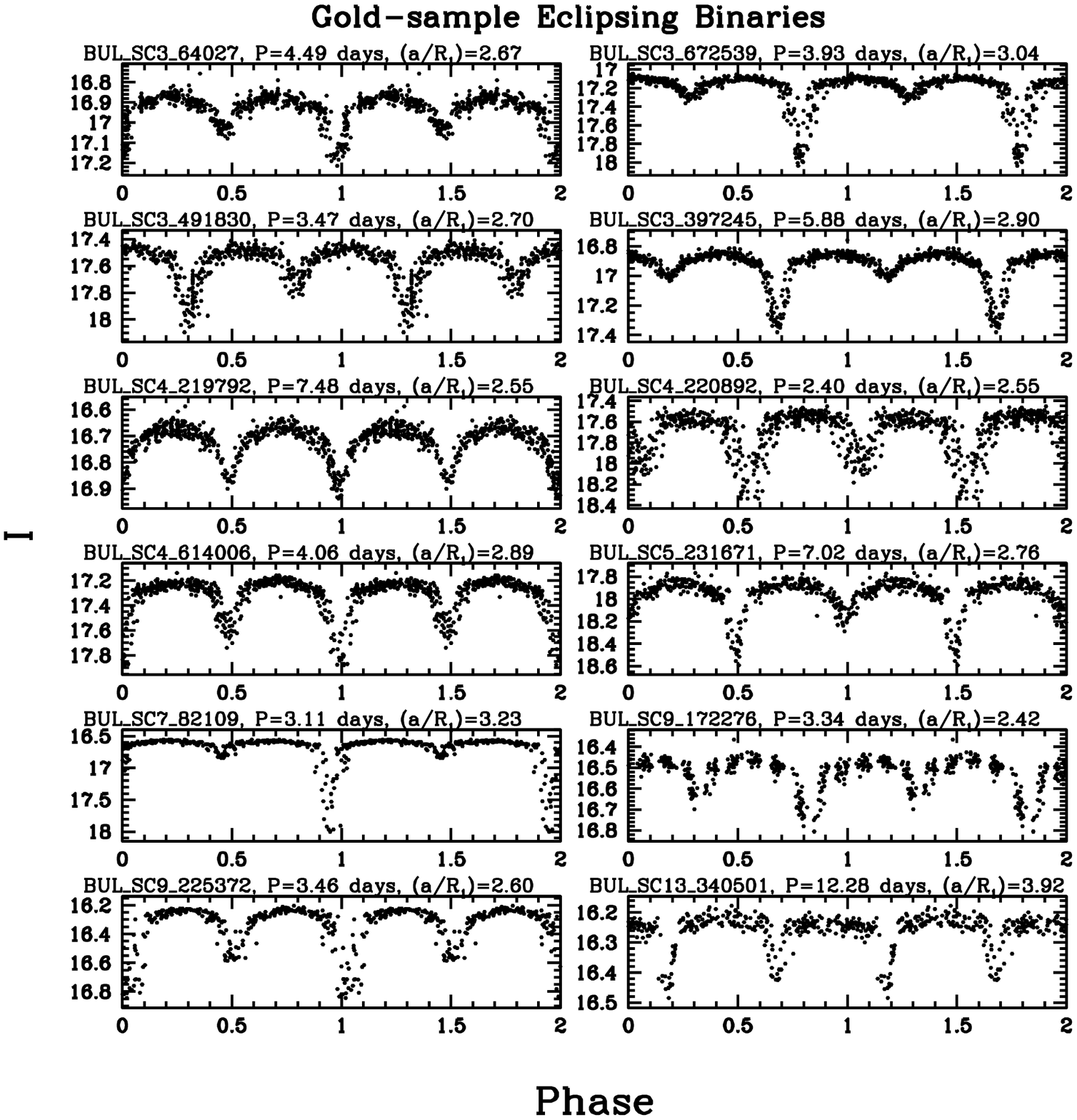}}
\vskip5pt
\FigCap{OGLE-II photometry (Udalski \etal 1997, Szyma�ski \etal 2005) 
for gold-sample eclipsing binary stars phase-folded over the periods
measured by Devor (2005). For each binary we state the OGLE-II
identification, as well as the period and ratio of semimajor axis to
primary radius measured by Devor (2005).}
\end{figure}
\begin{figure}[htb]
\centerline{\includegraphics[width=13cm]{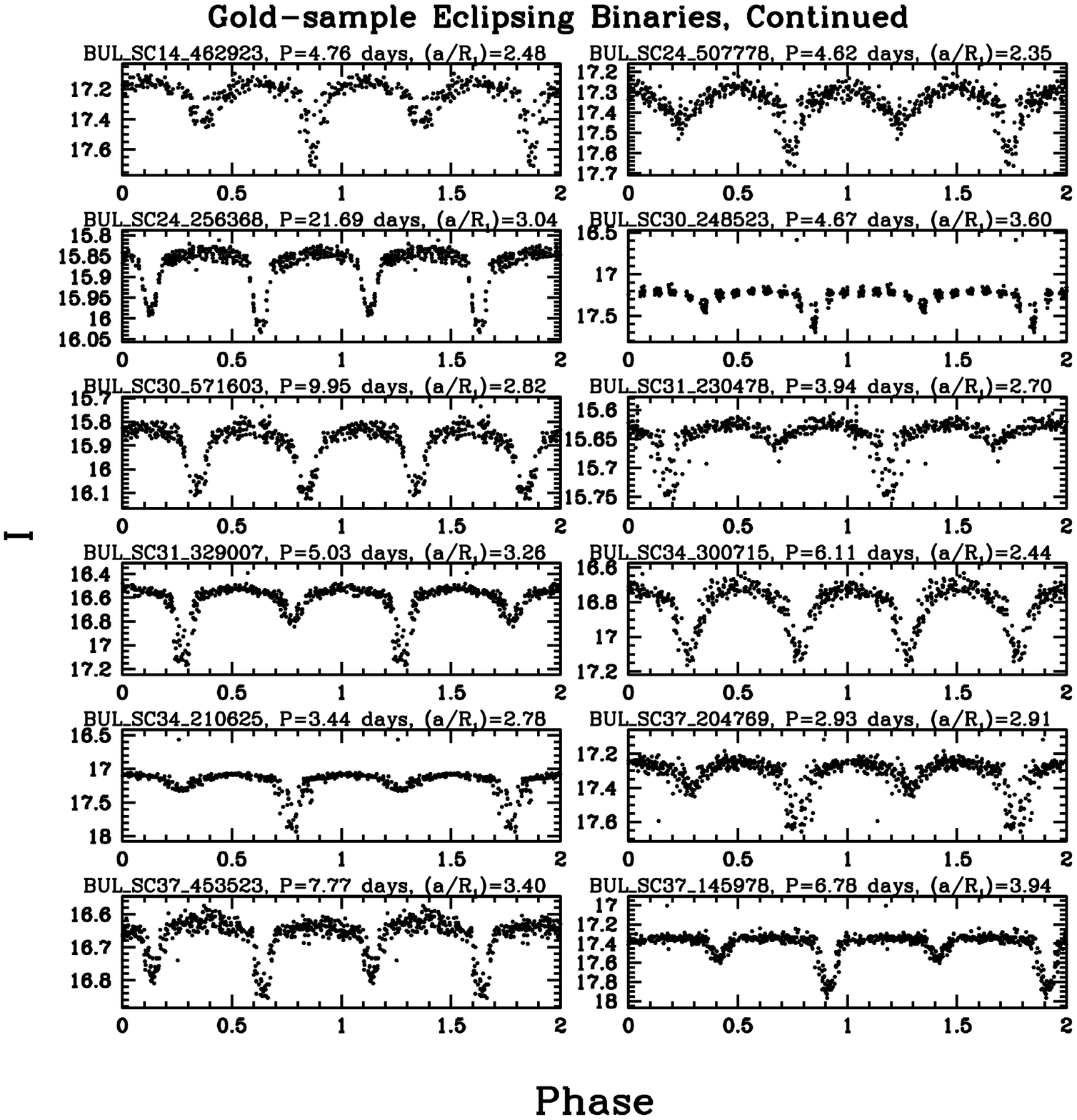}}
\FigCap{Same as in Fig.~3.}
\end{figure}

\begin{figure}[htb]
\centerline{\includegraphics[width=13cm, bb=0 035 547 573]{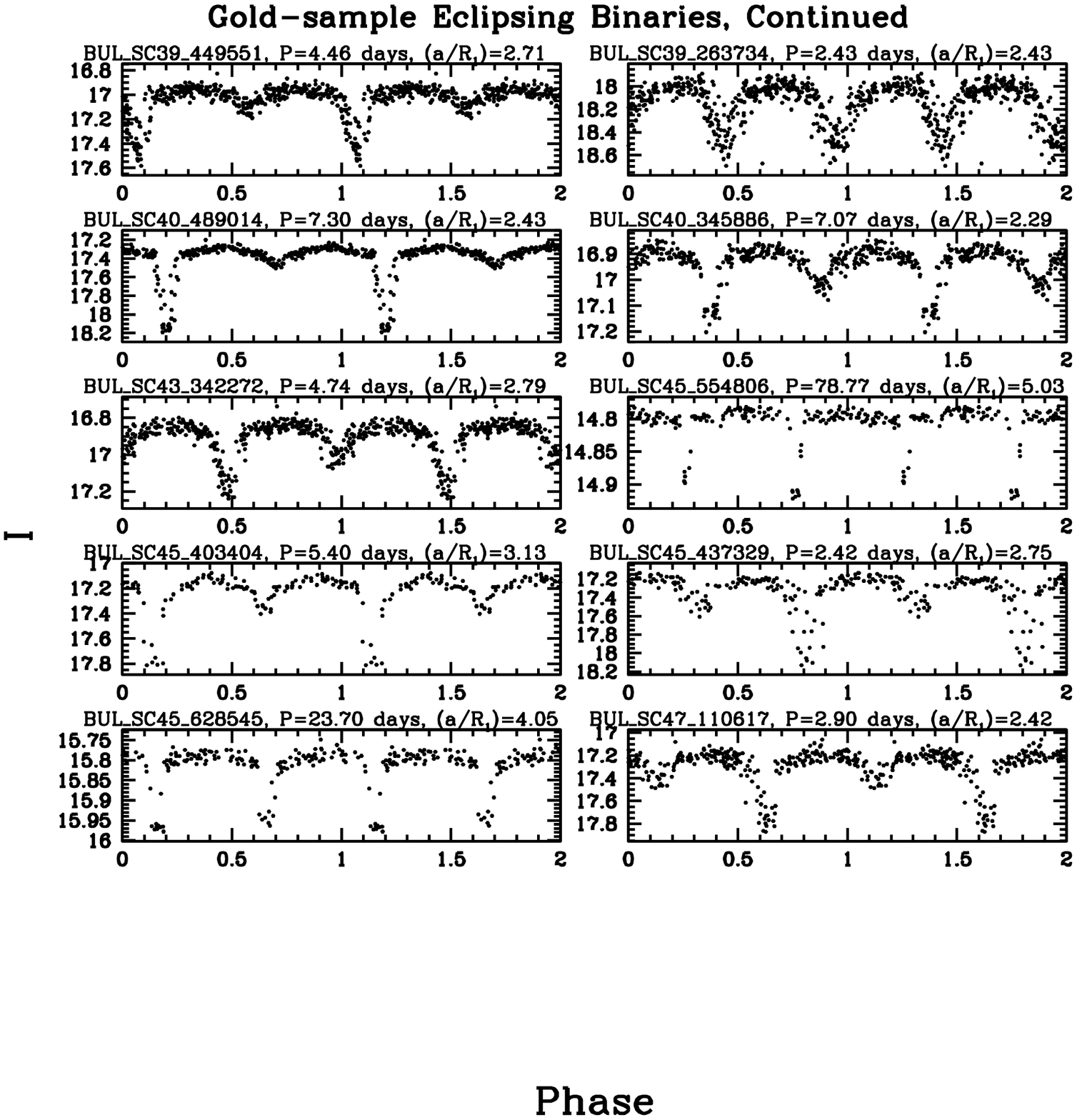}}
\FigCap{Same as in Fig.~3.}
\end{figure}

We compare our derived distribution of $R_2/R_1$ for both the gold and
silver samples, shown in panels (a) and (b) of Fig.~2, and we find that
these are not consistent with expectations: the catalog is likely missing
RG EB twins with both high and low radius ratios. The expected probability
density functions for $R_2/R_1$ are derived by sampling a scaled-solar
stellar track (Delahaye \etal 2010, see Section~3) in the evolution phase
during which $R\geq3~\RS$ but before the tip of the RG branch (mimicking
our selection for the gold sample). Due to the fact that the mass
difference for equal metallicity and co-eval stars on the RG branch is
small, it follows from the fuel-consumption theorem (Renzini and Buzzoni
1986) that the relative number counts are simply proportional to the
duration of specific phases of stellar evolution, in this case:
$$N(r)\dd r\propto\tau(r)\dd r.\eqno(4)$$
We randomly sample the stellar track in age (equivalent to sampling in
phase) $2\times10^6$ times, and in panel~(d) of Fig.~2 we show (in blue)
the resulting distribution of $R_2/R_1$. We also show a corrected predicted
probability density function (in red) that accounts for the empirical
period distribution of Duquennoy and Mayor (1991) given the assumption that
$M_1=M_2=1~\MS$, the requirement that the EBs be detached, and the
assumption of an eclipse probability $P(E)=(R_1+R_2)/a$. The ratio of
$R_2/R_1$ is predicted to increase smoothly, approximately doubling, over
the range $0\leq R_2/R_1\leq1$, with an excess at $R_2/R_1\approx0.45$ due
to the red giant branch bump (RGBB). Instead, both the gold and silver
sample peak at $R_2/R_1\approx0.75$, with a sharp drop-off at both ends.

We suggest two plausible reasons for the difference between the observed
and predicted distributions of $R_2/R_1$. At low values of $R_2/R_1$, the
photometric eclipse depth drops rapidly, and thus may evade detection by
the algorithm of Devor (2005). At high values of $R_2/R_1$, the eclipses
may become identical in shape, a solution that may be indirectly biased
against by the detection algorithm. We argue against the RC playing a
significant role in these biases. It is not included in our predicted
luminosity function, but as can be seen in the top left panel of Fig.~1,
very few of the EB twins (either gold or silver) are in the RC. There is a
sound theoretical reason to expect this. All RC stars will have previously
ascended the RG branch and reached very high values of $R/\RS$. Thus, if
they have a binary companion sufficiently close as to have detectable
eclipse once their their size shrinks to $R\approx 10~\RS$, they would have
significantly overflowed their Roche lobes, and thus likely ended up on a
section of the zero-age horizontal branch corresponding to higher
mass loss.

Due to the biases in the sample, we cannot assess whether our assay is
consistent with the claim that close binaries have a flat secondary mass
function, a result observed for various types of binary stars (Kuiper 1935,
Pinsonneault and Stanek 1996). If the detection efficiency is far below
100\%, then our estimate of 200 ``gold'' candidates waiting to be found in
OGLE-III data is an underestimate. We derived this number by taking the 34
candidates found in OGLE-II and scaling by the ratio ($\approx6$) of bulge
RR~Lyr found in OGLE-III (Soszy�ski \etal 2011) to that found in OGLE-II
(Collinge \etal 2006). However, if RG EB twins in OGLE-II had a low
detection efficiency, the benefit of using the higher-cadence,
longer-baseline, and more precise OGLE-III data will be substantially
higher than naively estimated.

\Section{Stellar Models}
We use the Yale Rotating Stellar Evolution Code (YREC, van Saders and
Pinsonneault 2012) for the theoretical predictions provided in this
work. The models are computed with diffusion. In order to match the
observed atmospheric metals-to-hydrogen ratio $Z/X=0.0229$ (Grevesse and
Sauval 1998), solar radius \RS, and solar luminosity \LS\ at $t=4.57$~Gyr,
the models used in this work have the solar composition set to
$(Z_{\odot},Y_{\odot})=(0.01875,0.27357)$, and mixing length parameter
$\alpha$ of 1.9449.

\Section{Early Scientific Payoffs of a Spectroscopic Survey of Detached Red Giant Eclipsing Binary Pairs}
\begin{figure}[b]
\vspace*{6pt}
\centerline{\includegraphics[width=9cm]{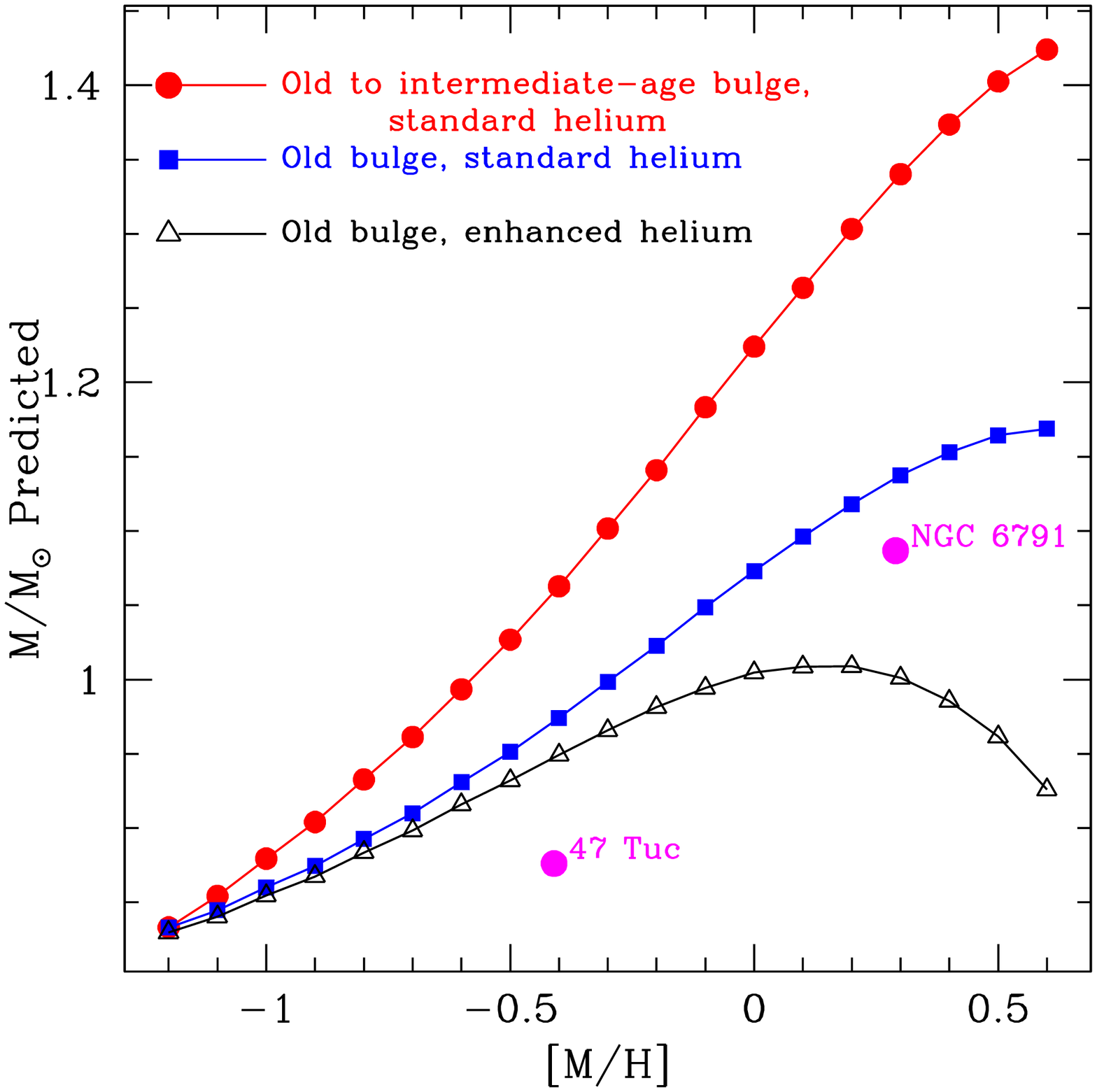}}
\vskip3pt
\FigCap{Different assessments of the age and helium abundance of 
the bulge predict sharply distinct, easily-measurable RG mass-metallicity
relationships. The old stellar population curves (blue, black) are
characterized by age-metallicity relations that are linear with
$\log(t/10)$ in the range $-1.2\leq[M/H]\leq+0.6$, $10\leq t/{\rm Gyr}
\leq12$, whereas the old-to-intermediate age curve (red) assumes $5\leq
t/{\rm Gyr}\leq12$ over the same metallicity range. The helium-enhanced
curve (black) assumes $\Delta Y/\Delta Z=3$. We show the approximate
turnoff mass (magenta) for the old, (relatively) metal-poor Galactic
globular cluster 47~Tuc (Thompson \etal 2010), and the approximate turnoff
mass for the old, metal-rich open cluster NGC~6791 (Brogaard \etal
2011). Both EBs shown in magenta are near the turnoff, and thus expected to
have masses slightly lower ($\Delta M\approx0.1~\MS$) than that on the RG
branch.}
\end{figure}
An early spin off of measurements of the Galactic bulge detached RG
eclipsing binary population would be either the validation or invalidation
of several predictions derived from the age-helium-metallicity relations
assumed and stated in the literature. The mass of RG stars is predicted to
have the following functional dependence on metallicity, age and initial
helium abundance:
$$\log\left(\frac{M}{\MS}\right)_{\rho=-3.0}=0.026+0.126~{\rm \left[\frac{M}{H}\right]}-0.276\log\left(\frac{t}{10}\right)-0.937(Y{-}0.27),\eqno(5)$$
and thus specific claims about the bulge age-helium-metallicity relation
derived from its CMD morphology predict equally specific mass--metallicity
relations for bulge RG stars. Fig.~6 shows that the mass--metallicity
relations predicted are very distinct. For example, Bensby \etal (2011)
assume isochrones with a standard helium-to-metals enrichment ratio and
find that stars with ${\rm [Fe/H]}=+0.35$ have an age of $\approx6$~Gyr. If
true, RGs at that metallicity should have masses of $\approx1.35$~\MS. If
this is the typical mass found for that metallicity it would confirm their
interpretation, and would invalidate the hypothesis that the entire bulge
stellar population formed by rapid gravitational in-fall, an invalidation
already suggested by dynamical investigations of metal-rich M-giants (Shen
\etal 2010, Kunder \etal 2011). Alternatively, if the bulge is enhanced in
helium, as argued based on measurements of the bulge red giant branch bump
(Nataf \etal 2011a,b) and the discrepancy between spectroscopic and
photometric turnoff ages (Nataf and Gould, 2011), then RG masses should be
significantly lower. If $(\Delta Y/\Delta Z)_{\rm bulge}=3$ and the
population is older than $t=10$~Gyr, no masses greater than $\approx1$~\MS\
should be measured on the RG branch.

The combination of RG mass and metallicity measurements with MSTO
single-star spectroscopic measurements is enough to determine both the
age-metallicity and helium-metallicity relationships of the bulge. Eq.~(5)
shows that at fixed metallicity, increased age has the same effect as
increased $Y$: both lead to decreased mass. This degeneracy has the
opposite angle to that found on the MSTO and subgiant branch (SGB):
increased $Y$ behaves similarly to decreased age, as both lead to higher
temperatures on the MSTO and lower surface gravities on the SGB (Nataf and
Gould 2011).

There could also be a strong metallicity dependence to the binary
fraction. Would the RG eclipsing binary metallicity-distribution function
(MDF) match that of the measured bulge RG MDF? If binaries are not a
representative sample of the underlying population, a significant fraction
of the bulge stellar population might not be directly probeable by this
method. This would be interesting in its own right, as it could constrain
models of how stars form in environments with different metallicity, an
issue recently brought into sharper focus by Conroy and van Dokkum
(2011). Moreover, there would still be value in measuring the age and
helium abundance of the bulge within the remaining metallicity range.

These stars would also be powerful dynamical probes. As these are bright
stars, their OGLE-III proper motions will be very precise. Their physical
radii would measure where these stars are located in the bulge relative to
the RC. This would give the distance, and would enable conversion of proper
motions into transverse velocities. The dynamics of the bulge are known to
be complex, with an X-shape at large separations ($Z\geq500$~pc) from the
plane recently discovered (Nataf \etal 2010, McWilliam and Zoccali 2010),
and correlations between kinematics and metallicity at all latitudes
(Babusiaux \etal 2010). A stellar sample with six measured kinematic phase
space coordinates is needed.

Mass loss along the RG branch is another scientific prospect. Brown
\etal (2011) found a significant number of low-mass white dwarfs without
binary companions, implying that a significant number of stars skip stages
of post-main-sequen\-ce stellar evolution, possibly due to enhanced mass loss
in metal-rich stars. This would be consistent with the spectroscopic study
of Rich \etal (2011), which showed that the ${\rm [Fe/H]}\approx+0.35$ peak
detected among bulge dwarf and SGB stars (Bensby \etal 2010, 2011) is not
present among bulge M-giants. Significant mass loss would manifest itself
as a much lower mass for more luminous stars, a characteristic that could
easily be measured in a sufficiently large survey.

\Section{Population Parameters and Observables}
The observable properties of a RG in an eclipsing binary pair are mass,
density, metallicity, and effective temperature. These four observables can
be matched to three theoretical quantities that effectively determine the
initial state of the star (mass, helium, and metallicity), plus the
evolutionary state at which we evaluate the stellar track. In principle,
this match between the number of independent parameters and measurements
enables complete determination of the age and helium abundance of each
detached RG eclipsing binary pair. For the purposes of this section, we
ignore theoretical uncertainties in the temperature and metallicity scale,
and discuss what can be done if one assumes accurate stellar models and
maximum-likelihood measurements and errors of the observables. We base our
parametrization on $\log\rho$ rather than $\log g$ because the
density, derived by plugging the EB light curve parameters into Kepler's 3rd
law, is typically measured with substantially higher accuracy than the
surface gravity.

Before moving forward, we must recognize that the accuracy of RG
temperature estimates as well as the interpretive power thereof is a matter
of ongoing controversy. There remains a $\approx100$~K uncertainty in the
temperature determination of stars (Casagrande \etal 2010), and this
uncertainty is comparable in size to the predicted effects of large age or
helium variations.  Moreover, for the sample of close binaries listed in
this work, $(R_{1,2}/a)\approx0.2$, and thus star-star interactions may
have a significant impact on the observed stellar properties (Chabrier
\etal 2007). Additionally, the convective efficiency assumed in this work,
parametrized by the mixing length, is calibrated on the Sun, and there is
no {\it a priori} reason why the efficiency should be the same in
RGs. However, there remains value in working out the standard theoretical
predictions, which may still be very effective for the most well-detached
systems. The prospects of calibrating the zero-point terms for RG
relationships are decent, due to the information accessible with missions
such as Kepler (\eg Hekker \etal 2010, 2011, Miglio \etal 2012). Additionally, 
whereas the interpretation of $T_{\rm eff}$ should be significantly
impacted by these concerns, that of $M/\MS$ should not -- the mass observed
in the RG phase is almost entirely a function of the main-sequence lifetime
for a given initial mass, metallicity and helium abundance. Any EB with
star-star interactions in the RG phase will not have experienced such
interactions during the main-sequence, as $R_{1,2}/a$ will have been
several times smaller.

Facilitating the task of parameter estimation is the fact that the
observable properties are predicted to be linear or nearly linear in the
parameter range of interest: stars of intermediate age and older. We use
our library of stellar tracks to derive relationships for the parameter
range $-3.5\leq\log(\rho/\rho_{\odot})\leq-2.5$, $-0.40\leq{\rm [M/H]} 
\leq0.40$, $5\leq t\leq15$~Gyr, $0.25\leq Y\leq0.40$:
$$D_i=\kappa_{ij}\eta_j+D_{i0}\eqno(6)$$
where  
$$D_i=
\begin{pmatrix} Y \\ 
\log{(t/10)} 
\end{pmatrix},\qquad
\eta_j= 
\begin{pmatrix} 
\log(T_{\rm eff})-3.65 \\ 
\rm{[M/H]} \\
\log(\rho/\rho_{\odot})+3\\ 
\log(M/\MS)  
\end{pmatrix},\eqno(7)$$
and 
$$\kappa_{ij}{=}\frac{\partial D_i}{\partial\eta_j}
{=}\begin{pmatrix}
  7.1208 &  0.28426 & -0.21878 & -0.50889 \\
-24.176  & -0.51046 &  0.7387  &  -1.8901  
\end{pmatrix}\!,\,\,
D_{i0}{=} 
\begin{pmatrix}
0.26516 \\
0.10957
\end{pmatrix}\!.\eqno(8)$$
The errors and covariance matrix in $D_i=(Y,\log(t/10))$ are given by 
$$\sigma_i=\sqrt{C_{ii}},\qquad
C_{ij}=\sum\limits_{m,n=1}^4\kappa_{im}\kappa_{jn}c_{mn}\eqno(9)$$
where $c_{ij}$ is the covariance matrix of the observables.

\begin{figure}[htb]
\includegraphics[width=12.3cm]{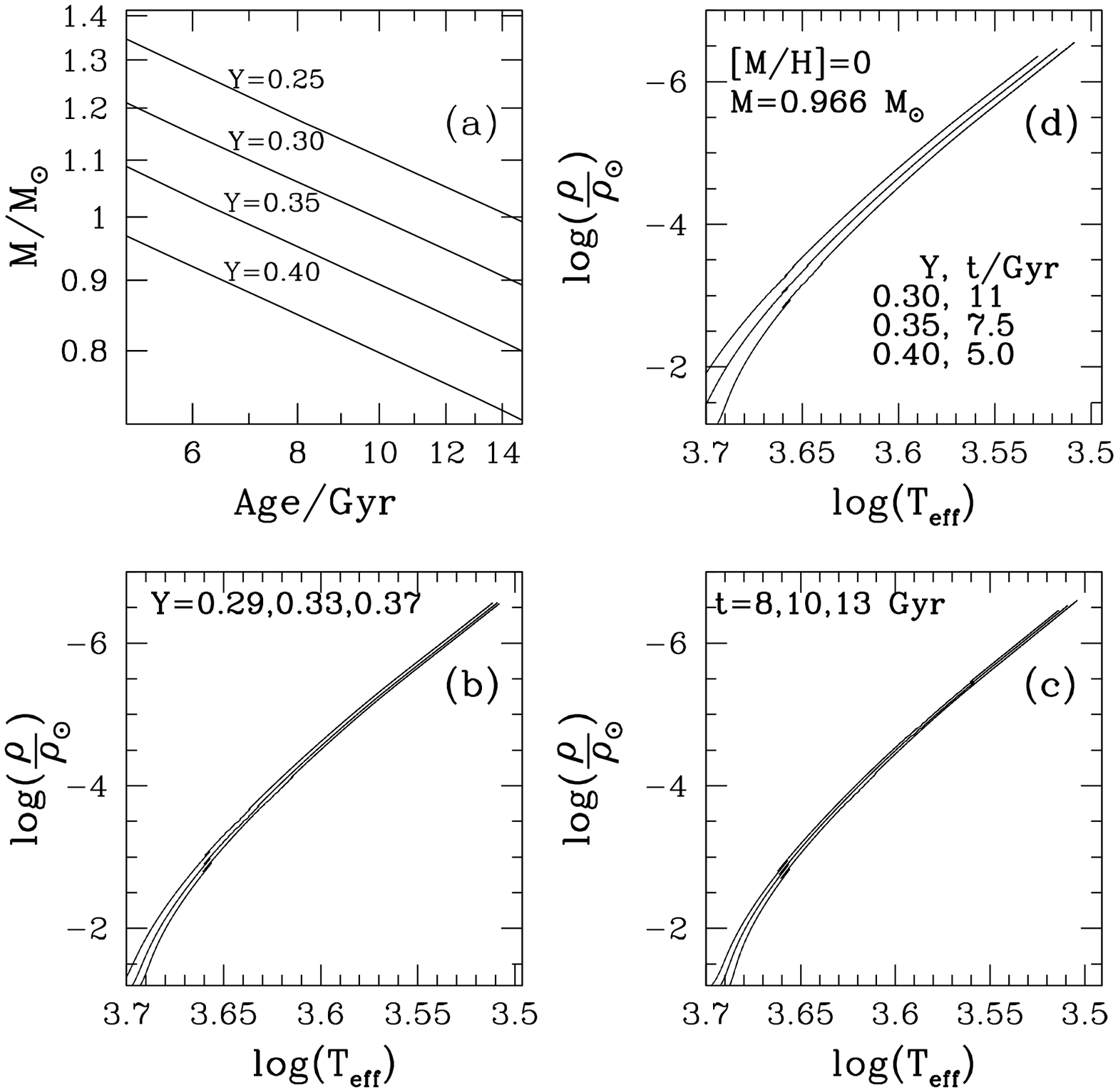}
\vspace*{6pt}
\FigCap{Predicted relationships between RG parameters and observables.
{\it Panel (a):} $M/\MS$ as a function of age for ${\rm [M/H]=0}$ tracks
varying in initial helium. Lower mass means higher age or higher helium.
{\it Panel (b):} RG tracks with ${\rm [M/H]=0}$, $t=11$~Gyr varying in
initial helium. Higher helium at fixed metallicity and density implies a
hotter RG branch. {\it Panel (c):} RG tracks with solar chemistry varying
in age. Higher age at fixed chemistry and density implies a colder RG
branch. {\it Panel (d):} these two effects add constructively when the mass
is fixed, because age goes down as helium abundance goes up.}
\end{figure}
Eq.~(8) again emphasizes the crucial role of a mass measurement. Note that a
change in any of the first three quantities ($\log T_{\rm eff}$, [Fe/H],
$\log\rho$) induces motions of $Y$ and $\log t$ in opposite directions,
while a change in $\log M$ induces motion in the same direction.
Therefore, without a tight mass measurement it is impossible to jointly
constrain the helium content and age of the star. This is illustrated
graphically in Figs.~7 and ~8.
\begin{figure}[htb]
\centerline{\includegraphics[width=10cm, bb=0 35 530 520]{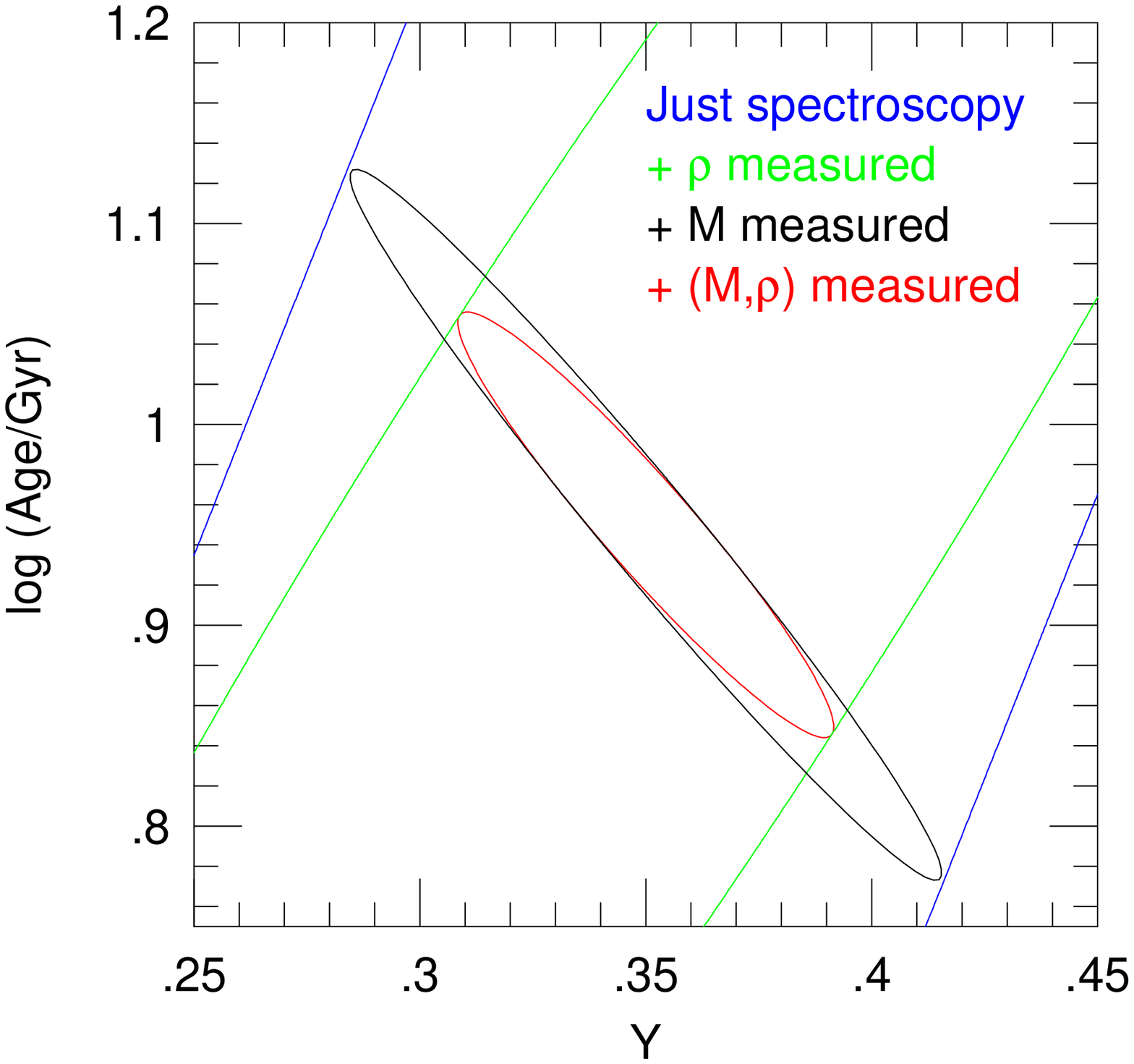}}
\FigCap{Predicted constraints on the age-helium plane from
spectroscopy only ($\log T_{\rm eff}$,[Fe/H],$\log g$) (blue), spectroscopy
plus density measurement (green), spectroscopy plus mass measurement
(black), and all measurements (red), assuming measured properties ($M/\MS$,
[M/H], $T_{\rm eff}$, $\log(\rho/\rho_{\odot})+3$) $= (0.91\pm0.01,
0.0\pm0.1, 4563\pm30, 0\pm0.01)$. With just spectroscopy, age and helium
are completely degenerate. Even precise measurement of the density only
narrows this degeneracy, it does not effectively break it. On the other
hand, a mass measurement generates an orthogonal constraint on this plane
and so does, by itself, break the degeneracy. Combining mass and density
measurements then further reduces the errors by a factor~1.6.}
\end{figure}

Fig.~7a shows the age-mass relation for RG stars at solar metallicity for
different helium abundances. Note that the curves are both straight and
equally spaced, \ie a linear relation. This implies that perfect mass and
metallicity measurements would yield a 1-dimensional linear constraint on
the $Y{-}\log(t/10)$ plane. In fact, while an essentially perfect mass
measurement is quite feasible (see below), this is not so for metallicity,
as such the 1-dimensional constraint would be a band rather than a line.

This band is almost perfectly orthogonal to the constraint that can be
obtained from spectroscopy ($\log T_{\rm eff}$, [Fe/H], $\log g$), which is
shown in Fig.~8. Like the mass/metallicity constraint, spectroscopy by
itself measures one fewer quantity than the number of model parameters
being constrained, and thus is represented by parallel-line error contours,
rather than closed contours. For purposes of this and subsequent plots, we
begin with the covariance matrix $c_{ij}=e_i e_j{\rm cor}_{ij}$, of a
typical RG star (from Alves-Brito \etal 2010)
$$e_j=\begin{pmatrix}
\sigma_{\log T_{\rm eff}} \\
\sigma_{\rm [M/H]}\\
\sigma_{\log g}  
\end{pmatrix}=\begin{pmatrix}
0.0044 \\
0.161 \\
0.184 
\end{pmatrix},\,\,\, \rm{cor}_{ij}=\begin{pmatrix}
1 & 0.36 & 0.39 \\
0.36 & 1 & -0.64 \\
0.39 & -0.64 & 1 \\
\end{pmatrix}\eqno(10)$$
which has been derived directly from the ensemble of line measurements,
using the method of Epstein \etal (2010). We then multiply the errors by a
factor 2/3, in recognition of the fact that the mass measurements will
require multiple epochs of high S/N, high-resolution spectra.

We now turn to the impact of a $\rho$ measurement, first combined only with
spectroscopy (Eq.~10) and then with a mass measurement as well. Unlike mass
and metallicity, the other three quantities that one can hope to directly
measure ($\log T_{\rm eff}$, $\log\rho$, $\log g$) all depend strongly on
phase of stellar evolution. In Fig.~7bc, we show these tracks on the $\log
T_{\rm eff}/\log\rho$ plane for various values of $Y$ and $\log(t)$
respectively. The main point to take away from these panels is that even if
the age (panel~b) or the helium content (panel~c) were known exactly, it
would be very difficult to distinguish the tracks from precise temperature
and density measurements. Only if the mass is measured, so these two
effects can be combined, do the tracks become well-separated (panel~d).

The impact of a density measurement (combined with spectroscopy) on the
age-helium plane is shown by the green curves in Fig.~8. Like the
spectroscopy-only constraint (blue), these appear to be parallel lines,
which is surprising given that the are four measurements $(\log T_{\rm
eff},{\rm [Fe/H]},\log g,\log\rho)$ to constrain four model parameters. In
fact, these contours are part of an ellipse which is extremely elongated in
the direction orthogonal to the mass constraint. The reason for this is
essentially the same as spectroscopy-only case: weak coupling of $\log M$
to $Y$ and $\log t$ {\it via} error propagation from the $\log g$
measurement. Fig.~8 also shows the impact of adding a mass measurement to
either of these cases. When all measurements are combined, the error
ellipse is highly elongated along the direction of the mass constraint,
with its width essentially determined by the metallicity error. Note the
$\log M$ and $\log\rho$ measurements have both been assumed to have errors
of 0.01~dex (2.3\%). However, this diagram would look almost exactly the
same if we had assumed zero errors for both quantities. Hence, even
conservative error bars are equivalent to perfect measurements for the
$\log M$ and $\log\rho$. We note that the detached RG EB twins OGLE SC10
137844, whose parameters are somewhat more difficult to measure due to the
fact they are in a highly eccentric ($e=0.31$) and long-period system
($P=372$~days), recently had their masses measured to $\approx0.8\%$ and
densities to $\approx4.5\%$ (Graczyk \etal 2012).

The actual spectroscopic measurements would be greatly constrained by the
eclipsing binary light curve. Aside from the precisely measured values of
$\log g$ obtained from $\log\rho$ and $M/\MS$, relative temperatures and
luminosities would be known to a high degree of accuracy from the
photometry alone, reducing the degrees of freedom allowed in the
spectroscopic fit. That the two stars are in a moderately close binary
would strongly suggest that they have identical metallicities. Further, the
ability to take spectra during eclipse, when only the larger and thus
brighter star would contribute to the spectrum, guarantees spectroscopic
parameters at least as good as those of a single RG. We note that for these
systems the eclipses often last several days, making the acquisition of
single-star spectra very feasible. We also comment on an important recent
finding. Gonzalez \etal (2011ab) used dereddened near-IR photometry of the
Galactic bulge and showed that the photometric color distribution of bulge
stars was in itself sufficient to reproduce the spectroscopic bulge
metallicity distribution function. This demonstrates that the bulge RG
temperatures are well-behaved.

\Section{Discussion}
\subsection{The Effect of Uncertainties in Stellar Evolution Models}
The stellar models that are used to interpret the measurements are
themselves a source of uncertainty. Heavy element diffusion, the value of
the mixing length, and angular momentum evolution are among the
uncertainties of stellar evolution. Whereas mass and metallicity
measurements should yield immediate and powerful constraining power,
temperature measurements will be more difficult to interpret due to their
greater theoretical uncertainties. This is due to the fact that stellar
models are typically calibrated with respect to the Sun, a main-sequence
star. Thus main-sequence behavior (\eg main-sequence lifetime as a
function of mass and composition) is better calibrated than
post-main-sequence behavior. Interpreting RG temperatures will therefore
require greater care than interpreting masses. There will be at least three
different means to constrain the impact of these uncertainties.

We ran several models with varied input parameters to gauge the effect of
two of the most significant uncertainties: the mixing length and diffusion.
Reducing the mixing length to 1.6449 from 1.9449, a shift comparable in
size to the empirical (Ferraro \etal 2006) and theoretical (Trampedach and
Stein 2011) determinations of the uncertainty, has the effect of increasing
the stellar lifetime at fixed mass, metallicity and helium by $\approx1\%$
-- the effect is negligible. For diffusion, we computed the same models
without diffusion, an exaggeration of the size of the error in diffusion,
and these yielded a stellar lifetime decrease of $\approx3\%$. Note that
turning off diffusion requires small changes in the metals abundance and
the mixing length to maintain consistency with the solar constraints, see
Table~1 of van Saders and Pinsonneault (2012). Both effects are smaller
than the statistical error resulting from a typical precision in the
metallicity of $\approx0.1$~dex. We also verified that the mass predictions
for RG stars in this work agree with those of the BaSTI (Pietrinferni \etal
2006) and Dartmouth (Dotter \etal 2008) stellar databases -- agreement is
to $\approx1\%$ on the lower RGB.

An excellent way to constrain the effect of theoretical uncertainties would
be to exploit the fact that the relative predictions of stellar evolution
models are more reliable than the absolute predictions. A reasonable choice
for an anchor point would be requiring the most metal-poor stars to have
the same age as the Galactic globular clusters (GCs), $t_{\rm
GC}=12.8\pm0.4$~Gyr (Mar{\'{\i}}n-Franch \etal 2009), and to have a
primordial helium abundance for those stars of $Y=0.249$ (Simha and
Steigman 2008). In principle, the metal-poor bulge may in fact be a little
older: bulge RR~Lyr stars are more metal-rich than those in globular
clusters (Kunder and Chaboyer 2008, Pietrukowicz \etal 2011), which might
imply that the most metal-poor bulge stars are too old to generate many RR
Lyrae stars. However, since the age difference between the GCs and the
universe is not large, such a difference would not significantly undermine
the use of the GCs as an age anchor point.

At Disk-like metallicities (\ie ${\rm [Fe/H]}\approx-0.3$), there is a
campaign to measure precise abundances of RG stars that have asteroseismic
measurements from the KEPLER and COROT satellites (Epstein \etal 2010,
NOAO-2011A-0510). Due to the fact that these are Disk stars, which are
expected to have younger ages and likely different helium-enrichment
patterns, we should expect that the $\log g-T_{\rm eff}-{\rm [M/H]}$ mapping
will not be identical to the mapping that can be obtained for bulge
stars. It will be interesting to see how they differ. The differences would
be combined with the predictions of Eq.~(8) to yield estimates of the
relative age and helium offsets between these two stellar populations.

At the metal-rich end, the open cluster NGC~6791 is a potent anchor. It has
a high metallicity ${\rm [Fe/H]}=+0.40$, it is the subject of detailed
asteroseismic study from Kepler photometry (Miglio \etal 2012), and has
three eclipsing binary pairs with measured masses and radii including two
near the main-sequence turnoff (Grundahl \etal 2008, Brogaard \etal
2011). An interesting finding of Miglio \etal (2012) is that the mass loss
for the metal-rich RG stars in NGC 6791 is not large:
$\Delta\overline{M}\approx0.09~\MS$. However, this finding is suggestive
rather than conclusive due to the use of first-order asteroseismic scaling
relations, which are still a matter of ongoing investigation. A direct mass
measurement of mass loss would be helpful, and this could be detected if it
were found that RG stars at the same metallicity have a
luminosity-dependent mass. With enough luck, one or several EBs at large
periods may even be found on the RC phase of stellar
evolution. Short-period EBs with RC members would not be adequate, as it is
likely that these detached EBs would not have been detached when the RC was
a much larger upper-RG branch star.

The combined use of these three anchors and the relative predictions of
stellar evolution could over determine the properties of bulge stars.
Because they span the entire metallicity range
$-1.2\leq\rm{[M/H]}\leq+0.60$ (Zoccali \etal 2008, Johnson \etal 2011, Hill
\etal 2011), any residual uncertainty could be used to place bounds on the
uncertainties in stellar models.

\subsection{Other Uncertainties}
There are several systematic effects that could pose challenges to any
survey of eclipsing binary pairs. We show, however, that these could either
be controlled or become investigative avenues in their own right.

The first is that blending can affect the light curve of bulge eclipsing
binaries, a systematic that is not generally a concern in the solar
neighborhood. The extra flux from a blend, $F_b$, would bias the
interpretation of the flux drop during the eclipse, and thus the value of
the derived stellar density. In Appendix, we derive that:
$$\Delta\ln\rho_1=-\frac{3}{4}\frac{F_b}{F_1}.\eqno(11)$$ 
It follows from Eqs.~(6) and (11) that a 10\% blend would significantly
affect the derived stellar parameter values. This could be controlled by
getting high-resolution images of the target in {\it JHK}. A significant
blend could also be estimated from its contribution to the spectra, as its
spectral lines would not share the $\approx100$~km/s orbital
velocities. Thus, it likely that the blending fraction ${F_b}/{F_1}$
could be measured to a precision of a few percent.

The second effect is the RGBB. This post-main-sequence phase of stellar
evolution, during which the star temporarily gets fainter, before getting
brighter again (Cassisi and Salaris 1997, Bjork and Chaboyer 2006, Nataf
\etal 2011b), breaks the injective mapping between density and temperature
at fixed mass, helium and metallicity. The resulting difference in
temperature can yield an offset in the derived values of $\delta Y\approx
0.007$ and $\delta\log{(t/10)}\approx0.02$. However, this would only occur
at the position of the RGBB, which can be easily estimated to a precision
of $\approx0.04$~dex in $\log g$ (\ie $\approx0.1$~mag) as its characteristic
luminosity and number counts are steep, empirically-calibrated functions of
metallicity (Nataf \etal 2011b). The lifetime of the RGBB for bulge stars
($\approx20$~Myr, Nataf \etal 2011a) does not contribute to the total age
uncertainty, as independent errors add in quadrature, and that of the
metallicity determination would dominate.

The third issue is that of RS Canum Venaticorum (RS~CVn) type stars (Eaton
and Hall 1979). These heavily spotted systems, common in binaries, would
have effective temperatures distinct from those predicted by naive
models. If close-in binaries were found to have lower temperatures at fixed
density and metallicity, this would demonstrate an effect due to tides and
angular momentum transfer. The superior photometry from OGLE-III, with its
greater cadence, longer baseline, and higher precision, should impose upper
bounds on the spot depth of eclipsing binary candidates, which could be
further investigated by measuring the calcium H+K emission in the
spectra. We note that identification of RS CVn could be a good way to
identify binaries that are not RG-EB twins, which would be a means of
expanding the sample.

Stellar rotation could also be a concern. While isolated solar-mass stars
tend to spin down over time, their cousins in binary systems are tidally
circularized after 7~Gyr for periods $P<15$~days and after 12~Gyr for
periods $P<20$~days (Mathieu \etal 2004). Although tidal synchronization is
not the same as circularization, the two are closely related. Thus, at
least the closer RG eclipsing systems were probably rotating faster than
the Sun ($P=25$~day) while they were on the main sequence. Sills \etal
(2000) showed that at fixed mass and composition, RG stars whose
progenitors were rotating with periods $P=8$~day are predicted to be
slightly colder than non-rotating RG stars yielding an estimated systematic
effect of $\Delta Y\approx0.01$. Since the RG eclipsing binaries are at
longer periods, the effect will be even smaller, but should nevertheless be
taken into account.

\Section{Conclusion}
We have demonstrated that there are observable, detached RG eclipsing
binary pairs in the Galactic bulge, and have constructed gold and silver
samples of 34 and 248 candidates respectively by imposing strict physical
consistency requirements on the total OGLE-II eclipsing binary sample of
Devor (2005). We have demonstrated that the derived masses, temperatures,
metallicities and densities assuming reasonable error estimates would give
powerful constraints on the formation and evolution of the Galaxy: both the
age and helium abundance would be tightly constrained, at every measured
metallicity. In addition to providing fundamental insights on Milky Way
assembly, a survey of these systems would have the potential to teach us
about the metallicity-dependence of the binary fraction, mass loss on the
RG branch and stellar models. Our count of 34-(282) such systems in the
OGLE-II eclipsing binary catalog leads us to estimate that at least
200-(1500) could be found in OGLE-III, just from the larger viewing area.

\Acknow{DMN was primarily supported by the NSERC grant PGSD3-403304-2011. 
DMN and AG were partially supported by the NSF grant AST-1103471. MHP was
partially supported by NASA grant NNX11AE04G. We thank Jan Skowron, Jason
Eastman, Francis Fekel and Martin Asplund for helpful discussions.}

\newpage
\centerline{\bf Appendix}
\vskip6pt
{\it Effect of Blending on Derived Eclipsing Binary Parameters}
\vskip12pt
We derive the error made in estimating the densities of the binary
components made by failing to detect (and so take into account) blending by
a third star within the PSF. For this purpose, we make the simplifying
assumptions of uniform surface brightness and highly unequal source sizes.
If these assumptions are relaxed, the derivation becomes much more
complicated, but the final results are similar.

Define
$$\tau_1\equiv\pi\frac{t_1}{P}\quad\tau_2\equiv\pi\frac{t_2}{P},\qquad z_1\equiv\frac{r_1}{a},\quad z_2\equiv\frac{r_2}{a}$$
where $t_1$ is the transit time, $t_2$ is the ingress time, $P$ is the
period, $a$ is the semimajor axis, and $r_1$ and $r_2$ are the two radii.
Note that $t_i$ and $P$ can be measured with essentially infinite
precision, because they depend only on timing data, and not on blending.
Hence, $\tau_i$ are also observables with quasi-infinite precision. Then,
for circular orbits,
$$\tau_1=z_1(1-\beta^2)^{1/2}\quad\tau_2 = z_2(1-\beta^2)^{-1/2}$$
where $\beta$ is the impact parameter. Hence, the precision of
$$\beta=\sqrt{1-\frac{\tau_1}{\tau_2}\frac{z_2}{z_1}}=\sqrt{1-\frac{t_1}{t_2}\frac{r_2}{r_1}}\qquad z_1z_2=\tau_1\tau_2$$
depends directly on how well the ratio of radii, $r_2/r_1$ can be measured.
The fluxes in and out of eclipse are related to the two surface brightnesses by
$$F_1=\pi S_1r_1^2, \qquad F_2=\pi[S_1(r_1^2-r_2^2)+S_2r_2^2]=F_1+\pi(S_2-S_1)r_2^2\qquad F_3 =F_1+\pi S_2 r_2^2$$
Hence,
$$\frac{F_2-F_1}{F_3-F_1}=1-\frac{S_1}{S_2}\Rightarrow\frac{S_1}{S_2}=\frac{F_3-F_2}{F_3-F_1}$$
$$\frac{F_3-F_1}{F_1}=\frac{S_2}{S_1}\frac{r_2^2}{r_1^2}\Rightarrow\frac{r_2^2}{r_1^2}=\frac{F_3-F_2}{F_1}\Rightarrow
\eta\equiv\frac{z_2}{z_1}=\frac{r_2}{r_1}=\sqrt{\frac{F_3-F_2}{F_1}}$$
where $\eta\equiv z_2/z_1$. The inverse density of the whole system is given
$$\frac{1}{\rho}=\frac{4\pi}{3}\frac{r_1^3+r_2^3}{M}=\frac{4\pi}{3}\frac{a^3}{M}(z_1^3+z_2^3)=\frac{G}{3\pi}P^2(z_1^3+z_2^3)=\frac{G}{3\pi}P^2z_1^3(1+\eta^3)$$
$$z_1=\sqrt\frac{\tau_1\tau_2}{\eta}\Rightarrow$$
$$\frac{1}{\rho}=\frac{G}{3\pi}P^2(\tau_1\tau_2)^{3/2}(\eta^{3/2}+\eta^{-3/2})=\frac{\pi^2G}{3}\frac{(t_1t_2)^{3/2}}{P}(\eta^{-3/2}+\eta^{3/2})$$
And, if the two masses are known (or assumed) to be equal:
$$\frac{1}{\rho_1}=\frac{2\pi^2G}{3}\frac{(t_1t_2)^{3/2}}{P}\eta^{3/2}=\frac{2\pi^2G}{3}\frac{(t_1 t_2)^{3/2}}{P}\biggl(\frac{F_1}{F_3-F_2}\biggr)^{3/4}$$

Now, the hardest thing to measure (assuming no blending) will be $t_2$
(ingress time). This is relatively short, so few data points. Plus, one
must actually understand limb darkening quite well to measure it. The only
quantity susceptible to blending is $\eta$. If the system is blended by
flux $F_b$, then all three fluxes $F_1,F_3,F_3$ are increased by this
same amount. In this case,
$$\frac{\dd\ln\eta^2}{\dd F_b}=-\frac{1}{F_1}\Rightarrow\frac{\dd\ln\rho_1}{\dd F_b}=-\frac{3}{4}\frac{1}{F_1}\Rightarrow\Delta\ln\rho_1=-\frac{3}{4}\frac{F_b}{F_1}.$$

Blending can be controlled in two complementary ways.  First,
high-resolution images on 8~m class telescopes should be able to detect all
sources within 2 FWHM of the target, down to 1--2\% in {\it
JHK}. Translating these fluxes into {\it I}-band will involve some error,
but should be statistically unbiased.

Second, any ambient source within the optical PSF with 1--2\% of target
flux can be detected as follows.  Since the periods are typically
$P=20$~days, the two sources will have relatively motion roughly 100~km/s,
while the line-widths should be smaller by a factor $r/2a$, even if the
stars are tidally locked. Therefore, the line systems of the two stars
should be quite well separated, enabling excellent empirical templates of
both from the ensemble of RV spectra. Then these templates can be
subtracted from each spectrum, shifted by the fit velocity. The sum of the
residuals of these fits will give a S/N $\sim$ few spectrum of any ambient
light, certainly enough to identify its source.

\renewcommand{\arraystretch}{0.9}
\setcounter{table}{0}
\MakeTableSep{rrccccccr}{12.5cm}{OGLE-II field and lightcurve
identification from Devor (2005), coordinates, $V-I$
colors, apparent magnitudes, relative radii and periods for the gold (first
part) and silver (second part) sample of identified RG eclipsing binary pairs}
{\hline
\douprule
Field & ID & RA & DEC & $(V-I)$ & $I$ & $R_1/a$ & $R_2/a$ & period [days]\\
\hline
 3 & 2325 & $17\uph53\upm15\zdot\ups31$ & $-30\arcd07\arcm57\zdot\arcs2$ & 2.398 & 16.977 & 0.375 & 0.263 & 4.488256 \\  
 3 & 2823 & $17\uph53\upm55\zdot\ups27$ & $-30\arcd05\arcm18\zdot\arcs9$ & 2.133 & 17.214 & 0.329 & 0.249 & 3.930081 \\  
 3 & 2929 & $17\uph53\upm45\zdot\ups96$ & $-30\arcd04\arcm33\zdot\arcs6$ & 2.245 & 17.604 & 0.370 & 0.276 & 3.471413 \\  
 3 & 7715 & $17\uph53\upm31\zdot\ups11$ & $-29\arcd34\arcm38\zdot\arcs7$ & 2.417 & 16.969 & 0.345 & 0.222 & 5.883186 \\ 
 4 &  830 & $17\uph54\upm33\zdot\ups33$ & $-30\arcd06\arcm18\zdot\arcs3$ & 2.477 & 16.781 & 0.392 & 0.305 & 7.477364 \\ 
 4 &  831 & $17\uph54\upm35\zdot\ups36$ & $-30\arcd06\arcm23\zdot\arcs2$ & 2.326 & 17.744 & 0.392 & 0.346 & 2.403032 \\ 
 4 & 1290 & $17\uph55\upm01\zdot\ups80$ & $-30\arcd03\arcm44\zdot\arcs6$ & 2.388 & 17.356 & 0.346 & 0.268 & 4.055058 \\ 
 5 & 1889 & $17\uph50\upm35\zdot\ups23$ & $-30\arcd09\arcm19\zdot\arcs5$ & 3.664 & 18.061 & 0.362 & 0.279 & 7.024886 \\ 
 7 & 1089 & $18\uph08\upm53\zdot\ups56$ & $-31\arcd54\arcm48\zdot\arcs1$ & 1.362 & 16.659 & 0.310 & 0.280 & 3.112654 \\ 
 9 &  132 & $18\uph24\upm17\zdot\ups25$ & $-22\arcd10\arcm52\zdot\arcs1$ & 1.989 & 16.577 & 0.414 & 0.276 & 3.336563 \\ 
 9 & 1502 & $18\uph24\upm07\zdot\ups57$ & $-21\arcd31\arcm22\zdot\arcs4$ & 1.787 & 16.390 & 0.384 & 0.306 & 3.458615 \\ 
13 & 1282 & $18\uph17\upm07\zdot\ups41$ & $-24\arcd03\arcm03\zdot\arcs3$ & 2.112 & 16.314 & 0.255 & 0.105 & 12.277672 \\ 
14 & 3912 & $17\uph47\upm11\zdot\ups83$ & $-22\arcd42\arcm21\zdot\arcs1$ & 2.288 & 17.316 & 0.403 & 0.269 & 4.758608 \\ 
24 & 1374 & $17\uph53\upm38\zdot\ups18$ & $-33\arcd01\arcm25\zdot\arcs4$ & 2.193 & 17.368 & 0.426 & 0.248 & 4.623804 \\ 
24 & 2461 & $17\uph53\upm03\zdot\ups90$ & $-32\arcd45\arcm12\zdot\arcs4$ & 2.423 & 16.021 & 0.329 & 0.190 & 21.687112 \\ 
30 & 1600 & $18\uph01\upm22\zdot\ups55$ & $-29\arcd03\arcm24\zdot\arcs0$ & 1.929 & 17.316 & 0.278 & 0.233 & 4.671058 \\ 
30 & 6658 & $18\uph01\upm27\zdot\ups97$ & $-28\arcd24\arcm07\zdot\arcs8$ & 1.946 & 15.935 & 0.354 & 0.272 & 9.952315 \\ 
31 &  631 & $18\uph02\upm14\zdot\ups25$ & $-28\arcd56\arcm11\zdot\arcs2$ & 1.945 & 16.331 & 0.370 & 0.289 & 3.937200 \\ 
31 & 2641 & $18\uph02\upm10\zdot\ups51$ & $-28\arcd30\arcm10\zdot\arcs2$ & 1.768 & 16.657 & 0.307 & 0.222 & 5.033917 \\ 
34 & 1369 & $17\uph58\upm11\zdot\ups35$ & $-29\arcd24\arcm54\zdot\arcs4$ & 2.610 & 17.027 & 0.410 & 0.251 & 6.113610 \\ 
34 & 6500 & $17\uph57\upm57\zdot\ups95$ & $-28\arcd48\arcm09\zdot\arcs3$ & 2.045 & 17.238 & 0.360 & 0.261 & 3.436820 \\ 
37 & 1846 & $17\uph52\upm30\zdot\ups15$ & $-30\arcd11\arcm44\zdot\arcs6$ & 2.331 & 17.367 & 0.344 & 0.311 & 2.934912 \\ 
37 & 6396 & $17\uph52\upm35\zdot\ups36$ & $-29\arcd40\arcm28\zdot\arcs6$ & 2.619 & 16.767 & 0.294 & 0.190 & 7.772554 \\ 
37 & 7093 & $17\uph52\upm15\zdot\ups51$ & $-29\arcd36\arcm14\zdot\arcs2$ & 3.147 & 17.517 & 0.254 & 0.232 & 6.784166 \\ 
39 & 1159 & $17\uph55\upm53\zdot\ups47$ & $-30\arcd02\arcm53\zdot\arcs9$ & 2.385 & 17.154 & 0.369 & 0.217 & 4.455969 \\ 
39 & 1604 & $17\uph55\upm31\zdot\ups28$ & $-29\arcd59\arcm03\zdot\arcs9$ & 2.574 & 18.199 & 0.411 & 0.320 & 2.430296 \\ 
40 &  310 & $17\uph51\upm33\zdot\ups38$ & $-33\arcd39\arcm04\zdot\arcs9$ & 2.682 & 17.419 & 0.412 & 0.294 & 7.297005 \\ 
40 &  621 & $17\uph51\upm18\zdot\ups82$ & $-33\arcd33\arcm59\zdot\arcs8$ & 2.459 & 17.004 & 0.436 & 0.204 & 7.069012 \\ 
43 & 3095 & $17\uph35\upm17\zdot\ups21$ & $-26\arcd47\arcm27\zdot\arcs8$ & 2.439 & 16.954 & 0.359 & 0.279 & 4.740494 \\ 
45 & 1064 & $18\uph04\upm04\zdot\ups03$ & $-30\arcd03\arcm52\zdot\arcs2$ & 2.048 & 14.894 & 0.199 & 0.058 & 78.765788 \\ 
45 & 1155 & $18\uph03\upm44\zdot\ups62$ & $-30\arcd01\arcm55\zdot\arcs3$ & 1.956 & 17.143 & 0.319 & 0.243 & 5.400987 \\ 
45 & 1522 & $18\uph03\upm42\zdot\ups13$ & $-29\arcd53\arcm05\zdot\arcs3$ & 1.640 & 17.350 & 0.364 & 0.358 & 2.424248 \\ 
45 & 2170 & $18\uph03\upm57\zdot\ups10$ & $-29\arcd39\arcm36\zdot\arcs7$ & 2.122 & 15.881 & 0.247 & 0.090 & 23.699590 \\ 
47 &  481 & $17\uph27\upm01\zdot\ups35$ & $-39\arcd50\arcm21\zdot\arcs5$ & 1.778 & 17.383 & 0.413 & 0.315 & 2.898527 \\ 
\hline
 1 & 1713 & $18\uph02\upm46\zdot\ups29$ & $-30\arcd04\arcm30\zdot\arcs5$ & 2.154 & 16.330 & 0.140 & 0.076 & 5.663370 \\ 
 1 & 2513 & $18\uph02\upm08\zdot\ups29$ & $-29\arcd53\arcm53\zdot\arcs5$ & 1.916 & 15.160 & 0.426 & 0.265 & 2.359235 \\ 
 1 & 2938 & $18\uph02\upm02\zdot\ups94$ & $-29\arcd48\arcm30\zdot\arcs5$ & 1.662 & 16.311 & 0.375 & 0.164 & 2.903954 \\ 
 1 & 3738 & $18\uph03\upm01\zdot\ups96$ & $-29\arcd40\arcm20\zdot\arcs3$ & 1.710 & 17.608 & 0.267 & 0.264 & 3.551679 \\ 
 1 & 3846 & $18\uph02\upm42\zdot\ups99$ & $-29\arcd38\arcm47\zdot\arcs2$ & 1.818 & 16.510 & 0.289 & 0.097 & 7.779557 \\ 
 1 & 3859 & $18\uph02\upm49\zdot\ups68$ & $-29\arcd38\arcm41\zdot\arcs0$ & 1.889 & 17.665 & 0.391 & 0.191 & 2.949637 \\ 
 2 &  892 & $18\uph04\upm35\zdot\ups40$ & $-29\arcd11\arcm58\zdot\arcs1$ & 1.764 & 16.882 & 0.416 & 0.145 & 4.541277 \\ 
 2 & 1301 & $18\uph04\upm24\zdot\ups76$ & $-29\arcd07\arcm16\zdot\arcs1$ & 1.628 & 16.505 & 0.253 & 0.079 & 6.464870 \\ 
 2 & 1800 & $18\uph04\upm09\zdot\ups41$ & $-29\arcd00\arcm56\zdot\arcs9$ & 1.506 & 17.095 & 0.300 & 0.199 & 2.339578 \\ 
 2 & 2542 & $18\uph04\upm54\zdot\ups38$ & $-28\arcd53\arcm55\zdot\arcs9$ & 1.395 & 15.973 & 0.360 & 0.208 & 3.085778 \\ 
\hline}
\setcounter{table}{0}
\MakeTableSepp{rrccccccr}{12.5cm}{Continued}
{\hline
\douprule
Field & ID & RA & DEC & $(V-I)$ & $I$ & $R_1/a$ & $R_2/a$ & period [days]\\
\hline
 2 & 3673 & $18\uph04\upm45\zdot\ups22$ & $-28\arcd42\arcm13\zdot\arcs0$ & 1.673 & 17.379 & 0.322 & 0.088 & 12.985678 \\ 
 2 & 3894 & $18\uph04\upm19\zdot\ups55$ & $-28\arcd39\arcm46\zdot\arcs3$ & 1.597 & 17.256 & 0.338 & 0.164 & 3.335057 \\ 
 2 & 4754 & $18\uph04\upm10\zdot\ups21$ & $-28\arcd30\arcm05\zdot\arcs0$ & 1.662 & 17.289 & 0.325 & 0.203 & 2.130568 \\ 
 3 & 1149 & $17\uph53\upm11\zdot\ups82$ & $-30\arcd16\arcm45\zdot\arcs6$ & 2.844 & 16.814 & 0.421 & 0.187 & 3.295681 \\ 
 3 & 1688 & $17\uph53\upm58\zdot\ups88$ & $-30\arcd13\arcm45\zdot\arcs6$ & 2.052 & 17.444 & 0.381 & 0.214 & 2.479170 \\ 
 3 & 2195 & $17\uph53\upm15\zdot\ups24$ & $-30\arcd08\arcm44\zdot\arcs6$ & 2.224 & 16.433 & 0.418 & 0.159 & 3.179790 \\ 
 3 & 3488 & $17\uph53\upm53\zdot\ups48$ & $-30\arcd01\arcm21\zdot\arcs8$ & 2.328 & 18.132 & 0.259 & 0.204 & 2.156910 \\ 
 3 & 3489 & $17\uph53\upm54\zdot\ups82$ & $-30\arcd00\arcm46\zdot\arcs7$ & 2.652 & 16.082 & 0.283 & 0.082 & 2.943854 \\ 
 3 & 3547 & $17\uph53\upm25\zdot\ups67$ & $-30\arcd00\arcm01\zdot\arcs5$ & 2.635 & 17.816 & 0.266 & 0.228 & 2.474109 \\ 
 3 & 3744 & $17\uph54\upm04\zdot\ups47$ & $-29\arcd59\arcm15\zdot\arcs8$ & 2.335 & 17.619 & 0.335 & 0.272 & 3.511041 \\ 
 3 & 6413 & $17\uph53\upm31\zdot\ups09$ & $-29\arcd41\arcm39\zdot\arcs3$ & 2.382 & 16.943 & 0.322 & 0.182 & 8.528036 \\ 
 3 & 8222 & $17\uph53\upm55\zdot\ups24$ & $-29\arcd31\arcm35\zdot\arcs3$ & 2.469 & 17.493 & 0.297 & 0.242 & 7.623718 \\ 
 3 & 8395 & $17\uph53\upm16\zdot\ups26$ & $-29\arcd30\arcm30\zdot\arcs0$ & 3.095 & 17.369 & 0.418 & 0.257 & 5.337132 \\ 
 4 & 1047 & $17\uph54\upm10\zdot\ups50$ & $-30\arcd04\arcm31\zdot\arcs6$ & 2.063 & 16.928 & 0.316 & 0.176 & 2.075691 \\ 
 4 & 1120 & $17\uph54\upm44\zdot\ups27$ & $-30\arcd04\arcm11\zdot\arcs8$ & 2.206 & 16.953 & 0.448 & 0.227 & 7.343176 \\ 
 4 & 1224 & $17\uph54\upm21\zdot\ups76$ & $-30\arcd03\arcm07\zdot\arcs7$ & 2.226 & 17.284 & 0.362 & 0.216 & 5.331105 \\ 
 4 & 1289 & $17\uph54\upm57\zdot\ups96$ & $-30\arcd03\arcm31\zdot\arcs4$ & 2.317 & 18.168 & 0.267 & 0.200 & 3.543221 \\ 
 4 & 3707 & $17\uph54\upm58\zdot\ups48$ & $-29\arcd49\arcm18\zdot\arcs1$ & 1.824 & 17.100 & 0.300 & 0.198 & 2.381842 \\ 
 4 & 4049 & $17\uph54\upm41\zdot\ups33$ & $-29\arcd47\arcm16\zdot\arcs7$ & 2.424 & 16.459 & 0.119 & 0.103 & 3.842130 \\ 
 4 & 4449 & $17\uph54\upm07\zdot\ups59$ & $-29\arcd44\arcm58\zdot\arcs6$ & 2.321 & 15.552 & 0.572 & 0.129 & 2.269646 \\ 
 4 & 4564 & $17\uph54\upm57\zdot\ups61$ & $-29\arcd45\arcm02\zdot\arcs3$ & 2.502 & 17.463 & 0.328 & 0.211 & 3.350001 \\ 
 4 & 5392 & $17\uph54\upm17\zdot\ups17$ & $-29\arcd39\arcm20\zdot\arcs2$ & 2.294 & 16.289 & 0.162 & 0.105 & 2.820490 \\ 
 4 & 6839 & $17\uph54\upm39\zdot\ups93$ & $-29\arcd30\arcm30\zdot\arcs7$ & 2.033 & 17.402 & 0.319 & 0.227 & 2.379704 \\ 
 4 & 6905 & $17\uph54\upm18\zdot\ups89$ & $-29\arcd29\arcm39\zdot\arcs7$ & 2.137 & 18.454 & 0.305 & 0.296 & 3.514754 \\ 
 4 & 7346 & $17\uph54\upm12\zdot\ups23$ & $-29\arcd27\arcm05\zdot\arcs6$ & 2.371 & 17.937 & 0.323 & 0.247 & 2.697201 \\ 
 4 & 8604 & $17\uph54\upm10\zdot\ups56$ & $-29\arcd18\arcm18\zdot\arcs2$ & 2.488 & 18.280 & 0.201 & 0.128 & 4.274822 \\ 
 5 & 5124 & $17\uph49\upm58\zdot\ups22$ & $-29\arcd44\arcm04\zdot\arcs5$ & 4.193 & 17.418 & 0.169 & 0.125 & 5.659006 \\ 
 6 &  358 & $18\uph07\upm46\zdot\ups98$ & $-32\arcd29\arcm20\zdot\arcs1$ & 1.739 & 15.439 & 0.246 & 0.064 & 3.476010 \\ 
 6 &  598 & $18\uph08\upm28\zdot\ups11$ & $-32\arcd24\arcm58\zdot\arcs7$ & 1.731 & 18.033 & 0.326 & 0.158 & 2.457010 \\ 
 6 & 1214 & $18\uph08\upm34\zdot\ups60$ & $-32\arcd12\arcm28\zdot\arcs4$ & 1.683 & 16.958 & 0.431 & 0.242 & 5.688314 \\ 
 6 & 1517 & $18\uph07\upm56\zdot\ups87$ & $-32\arcd06\arcm12\zdot\arcs5$ & 1.335 & 17.511 & 0.102 & 0.078 & 6.417924 \\ 
 7 &  299 & $18\uph08\upm51\zdot\ups59$ & $-32\arcd23\arcm13\zdot\arcs2$ & 1.504 & 18.265 & 0.087 & 0.078 & 4.048338 \\ 
 7 &  605 & $18\uph09\upm14\zdot\ups45$ & $-32\arcd10\arcm42\zdot\arcs7$ & 1.865 & 17.022 & 0.383 & 0.324 & 4.259042 \\ 
 7 & 1494 & $18\uph08\upm52\zdot\ups25$ & $-31\arcd42\arcm44\zdot\arcs0$ & 1.410 & 17.626 & 0.134 & 0.113 & 3.341994 \\ 
 8 &  430 & $18\uph23\upm30\zdot\ups17$ & $-22\arcd03\arcm23\zdot\arcs6$ & 2.026 & 16.850 & 0.416 & 0.076 & 3.499018 \\ 
 8 &  829 & $18\uph23\upm23\zdot\ups13$ & $-21\arcd54\arcm38\zdot\arcs5$ & 1.969 & 17.260 & 0.239 & 0.198 & 4.084482 \\ 
10 & 1924 & $18\uph20\upm06\zdot\ups01$ & $-22\arcd09\arcm22\zdot\arcs3$ & 1.951 & 17.836 & 0.215 & 0.111 & 4.128718 \\ 
10 & 2053 & $18\uph20\upm00\zdot\ups77$ & $-22\arcd07\arcm06\zdot\arcs5$ & 1.778 & 16.923 & 0.311 & 0.286 & 3.729120 \\ 
11 & 1231 & $18\uph20\upm54\zdot\ups55$ & $-22\arcd21\arcm07\zdot\arcs1$ & 1.910 & 16.945 & 0.349 & 0.245 & 3.733986 \\ 
11 & 1934 & $18\uph21\upm22\zdot\ups82$ & $-22\arcd03\arcm02\zdot\arcs7$ & 2.119 & 16.015 & 0.289 & 0.106 & 11.554690 \\ 
12 & 1664 & $18\uph16\upm26\zdot\ups54$ & $-24\arcd00\arcm08\zdot\arcs6$ & 2.071 & 17.492 & 0.180 & 0.121 & 2.812648 \\ 
12 & 2249 & $18\uph16\upm22\zdot\ups47$ & $-23\arcd50\arcm50\zdot\arcs8$ & 1.909 & 17.937 & 0.177 & 0.109 & 4.133296 \\ 
12 & 3118 & $18\uph15\upm57\zdot\ups17$ & $-23\arcd35\arcm39\zdot\arcs6$ & 2.105 & 17.756 & 0.215 & 0.111 & 2.731754 \\ 
13 &  515 & $18\uph16\upm54\zdot\ups03$ & $-24\arcd16\arcm52\zdot\arcs7$ & 2.179 & 17.315 & 0.209 & 0.198 & 2.280115 \\ 
13 &  643 & $18\uph17\upm21\zdot\ups43$ & $-24\arcd15\arcm08\zdot\arcs1$ & 2.228 & 18.266 & 0.429 & 0.229 & 2.543559 \\ 
13 &  748 & $18\uph16\upm47\zdot\ups11$ & $-24\arcd11\arcm53\zdot\arcs1$ & 2.477 & 17.513 & 0.184 & 0.085 & 3.086584 \\ 
14 &  748 & $17\uph46\upm47\zdot\ups93$ & $-23\arcd24\arcm30\zdot\arcs5$ & 2.054 & 16.809 & 0.270 & 0.116 & 3.205315 \\ 
\hline}
\setcounter{table}{0}
\MakeTableSepp{rrccccccr}{12.5cm}{Continued}
{\hline
\douprule
Field & ID & RA & DEC & $(V-I)$ & $I$ & $R_1/a$ & $R_2/a$ & period [days]\\
\hline
14 &  842 & $17\uph46\upm34\zdot\ups66$ & $-23\arcd22\arcm50\zdot\arcs8$ & 2.108 & 15.063 & 0.281 & 0.123 & 4.724715 \\ 
14 & 1246 & $17\uph47\upm16\zdot\ups93$ & $-23\arcd16\arcm39\zdot\arcs8$ & 2.224 & 16.485 & 0.256 & 0.092 & 10.525937 \\ 
14 & 1360 & $17\uph46\upm51\zdot\ups72$ & $-23\arcd14\arcm51\zdot\arcs2$ & 2.042 & 18.241 & 0.278 & 0.107 & 5.335997 \\ 
14 & 2296 & $17\uph47\upm19\zdot\ups49$ & $-23\arcd03\arcm42\zdot\arcs5$ & 2.428 & 15.460 & 0.240 & 0.056 & 2.429135 \\ 
14 & 3223 & $17\uph47\upm01\zdot\ups51$ & $-22\arcd53\arcm10\zdot\arcs2$ & 2.043 & 17.372 & 0.307 & 0.168 & 2.697653 \\ 
14 & 3755 & $17\uph46\upm38\zdot\ups18$ & $-22\arcd43\arcm40\zdot\arcs9$ & 1.996 & 17.495 & 0.317 & 0.237 & 2.063695 \\ 
15 &  529 & $17\uph47\upm50\zdot\ups68$ & $-23\arcd25\arcm54\zdot\arcs5$ & 2.122 & 17.324 & 0.371 & 0.172 & 4.172854 \\ 
15 &  631 & $17\uph48\upm16\zdot\ups06$ & $-23\arcd25\arcm22\zdot\arcs9$ & 2.125 & 17.695 & 0.285 & 0.228 & 2.886554 \\ 
15 &  851 & $17\uph48\upm12\zdot\ups09$ & $-23\arcd21\arcm51\zdot\arcs4$ & 2.277 & 17.754 & 0.385 & 0.191 & 3.268354 \\ 
15 & 1341 & $17\uph47\upm50\zdot\ups79$ & $-23\arcd15\arcm13\zdot\arcs1$ & 2.424 & 17.552 & 0.198 & 0.116 & 2.767438 \\ 
15 & 2256 & $17\uph48\upm13\zdot\ups84$ & $-23\arcd02\arcm45\zdot\arcs3$ & 2.297 & 17.944 & 0.376 & 0.270 & 2.562922 \\ 
16 &  147 & $18\uph10\upm03\zdot\ups80$ & $-26\arcd44\arcm29\zdot\arcs4$ & 2.035 & 18.238 & 0.190 & 0.094 & 7.611000 \\ 
16 &  443 & $18\uph09\upm55\zdot\ups54$ & $-26\arcd40\arcm37\zdot\arcs0$ & 2.073 & 17.613 & 0.338 & 0.229 & 2.515553 \\ 
16 &  722 & $18\uph09\upm58\zdot\ups67$ & $-26\arcd36\arcm52\zdot\arcs8$ & 2.165 & 17.419 & 0.406 & 0.268 & 5.064416 \\ 
16 & 2304 & $18\uph09\upm58\zdot\ups50$ & $-26\arcd19\arcm18\zdot\arcs5$ & 1.821 & 16.691 & 0.356 & 0.137 & 3.572162 \\ 
16 & 3066 & $18\uph09\upm45\zdot\ups75$ & $-26\arcd10\arcm33\zdot\arcs8$ & 1.977 & 17.956 & 0.287 & 0.179 & 4.327591 \\ 
17 &  417 & $18\uph11\upm10\zdot\ups05$ & $-26\arcd34\arcm35\zdot\arcs0$ & 1.639 & 16.783 & 0.369 & 0.235 & 2.958341 \\ 
17 & 1576 & $18\uph10\upm51\zdot\ups97$ & $-26\arcd21\arcm58\zdot\arcs3$ & 1.902 & 17.489 & 0.418 & 0.125 & 3.480961 \\ 
17 & 1577 & $18\uph10\upm58\zdot\ups42$ & $-26\arcd22\arcm06\zdot\arcs7$ & 2.019 & 16.847 & 0.181 & 0.130 & 20.532350 \\ 
17 & 2366 & $18\uph11\upm07\zdot\ups56$ & $-26\arcd11\arcm28\zdot\arcs4$ & 1.663 & 17.709 & 0.150 & 0.136 & 2.771794 \\ 
17 & 4107 & $18\uph11\upm07\zdot\ups22$ & $-25\arcd49\arcm57\zdot\arcs8$ & 1.863 & 16.939 & 0.367 & 0.153 & 3.584156 \\ 
18 &  667 & $18\uph06\upm37\zdot\ups36$ & $-27\arcd33\arcm42\zdot\arcs9$ & 2.116 & 16.079 & 0.172 & 0.074 & 9.504011 \\ 
18 & 1201 & $18\uph06\upm42\zdot\ups52$ & $-27\arcd28\arcm03\zdot\arcs3$ & 1.553 & 17.341 & 0.274 & 0.207 & 3.936500 \\ 
18 & 2492 & $18\uph06\upm38\zdot\ups20$ & $-27\arcd17\arcm03\zdot\arcs5$ & 1.509 & 17.265 & 0.254 & 0.101 & 3.187028 \\ 
18 & 3232 & $18\uph07\upm07\zdot\ups99$ & $-27\arcd09\arcm23\zdot\arcs8$ & 1.668 & 16.559 & 0.207 & 0.117 & 2.345478 \\ 
18 & 3834 & $18\uph06\upm33\zdot\ups83$ & $-27\arcd03\arcm35\zdot\arcs0$ & 1.649 & 17.338 & 0.153 & 0.082 & 4.064586 \\ 
18 & 3942 & $18\uph06\upm43\zdot\ups70$ & $-27\arcd02\arcm38\zdot\arcs1$ & 1.595 & 16.204 & 0.259 & 0.242 & 4.271732 \\ 
18 & 4093 & $18\uph07\upm25\zdot\ups56$ & $-27\arcd01\arcm43\zdot\arcs0$ & 1.854 & 17.096 & 0.332 & 0.229 & 3.373929 \\ 
18 & 4246 & $18\uph06\upm59\zdot\ups39$ & $-27\arcd00\arcm14\zdot\arcs5$ & 1.552 & 17.062 & 0.380 & 0.267 & 5.194025 \\ 
18 & 4368 & $18\uph07\upm03\zdot\ups12$ & $-26\arcd59\arcm28\zdot\arcs2$ & 1.596 & 17.437 & 0.213 & 0.141 & 2.582868 \\ 
19 & 3463 & $18\uph07\upm44\zdot\ups88$ & $-27\arcd02\arcm14\zdot\arcs3$ & 2.022 & 17.794 & 0.310 & 0.207 & 2.165127 \\ 
19 & 4115 & $18\uph08\upm31\zdot\ups28$ & $-26\arcd56\arcm27\zdot\arcs2$ & 1.699 & 17.174 & 0.325 & 0.244 & 2.022135 \\ 
20 &  508 & $17\uph59\upm17\zdot\ups21$ & $-29\arcd15\arcm46\zdot\arcs0$ & 1.784 & 16.925 & 0.346 & 0.246 & 6.437626 \\ 
20 &  625 & $17\uph59\upm11\zdot\ups17$ & $-29\arcd14\arcm26\zdot\arcs7$ & 1.713 & 16.759 & 0.267 & 0.160 & 2.156676 \\ 
20 &  726 & $17\uph59\upm12\zdot\ups62$ & $-29\arcd13\arcm58\zdot\arcs6$ & 1.620 & 17.047 & 0.359 & 0.323 & 3.220336 \\ 
20 &  764 & $17\uph59\upm37\zdot\ups56$ & $-29\arcd13\arcm20\zdot\arcs1$ & 1.788 & 17.347 & 0.308 & 0.134 & 4.828995 \\ 
20 & 1218 & $17\uph59\upm32\zdot\ups08$ & $-29\arcd09\arcm34\zdot\arcs2$ & 1.773 & 17.572 & 0.303 & 0.250 & 4.204415 \\ 
20 & 3228 & $17\uph59\upm39\zdot\ups47$ & $-28\arcd49\arcm45\zdot\arcs6$ & 2.033 & 15.511 & 0.301 & 0.084 & 2.761122 \\ 
20 & 3507 & $17\uph59\upm23\zdot\ups84$ & $-28\arcd47\arcm26\zdot\arcs0$ & 1.637 & 17.043 & 0.358 & 0.250 & 2.431410 \\ 
20 & 3703 & $17\uph59\upm09\zdot\ups40$ & $-28\arcd45\arcm47\zdot\arcs1$ & 2.053 & 18.174 & 0.200 & 0.182 & 7.134314 \\ 
20 & 3831 & $17\uph59\upm26\zdot\ups04$ & $-28\arcd44\arcm13\zdot\arcs9$ & 1.958 & 16.494 & 0.121 & 0.071 & 3.034756 \\ 
20 & 4260 & $17\uph59\upm36\zdot\ups54$ & $-28\arcd40\arcm51\zdot\arcs1$ & 1.833 & 17.474 & 0.214 & 0.107 & 4.127868 \\ 
20 & 5428 & $17\uph58\upm56\zdot\ups59$ & $-28\arcd29\arcm28\zdot\arcs5$ & 2.034 & 17.265 & 0.269 & 0.198 & 2.185926 \\ 
21 &  564 & $18\uph00\upm46\zdot\ups24$ & $-29\arcd15\arcm53\zdot\arcs4$ & 1.970 & 17.238 & 0.365 & 0.255 & 2.753914 \\ 
21 & 1563 & $18\uph00\upm34\zdot\ups98$ & $-29\arcd07\arcm43\zdot\arcs6$ & 1.954 & 15.966 & 0.184 & 0.032 & 28.799444 \\ 
21 & 1588 & $18\uph00\upm43\zdot\ups62$ & $-29\arcd08\arcm08\zdot\arcs8$ & 1.733 & 15.282 & 0.312 & 0.224 & 3.740300 \\ 
21 & 2360 & $18\uph00\upm22\zdot\ups92$ & $-29\arcd01\arcm28\zdot\arcs2$ & 1.543 & 16.283 & 0.520 & 0.176 & 3.589142 \\ 
21 & 2601 & $18\uph00\upm06\zdot\ups28$ & $-29\arcd00\arcm11\zdot\arcs0$ & 1.579 & 17.763 & 0.230 & 0.132 & 3.180098 \\ 
\hline}
\setcounter{table}{0}
\MakeTableSepp{rrccccccr}{12.5cm}{Continued}
{\hline
\douprule
Field & ID & RA & DEC & $(V-I)$ & $I$ & $R_1/a$ & $R_2/a$ & period [days]\\
\hline
21 & 2750 & $18\uph00\upm35\zdot\ups67$ & $-28\arcd58\arcm56\zdot\arcs2$ & 1.534 & 17.480 & 0.325 & 0.255 & 2.220600 \\ 
21 & 3596 & $18\uph00\upm09\zdot\ups93$ & $-28\arcd52\arcm40\zdot\arcs9$ & 2.038 & 18.247 & 0.360 & 0.334 & 4.252551 \\ 
21 & 4732 & $18\uph00\upm33\zdot\ups33$ & $-28\arcd45\arcm29\zdot\arcs7$ & 2.181 & 17.291 & 0.223 & 0.062 & 2.871256 \\ 
21 & 6233 & $18\uph00\upm03\zdot\ups29$ & $-28\arcd33\arcm32\zdot\arcs7$ & 2.488 & 16.801 & 0.415 & 0.141 & 2.923352 \\ 
21 & 6795 & $18\uph00\upm09\zdot\ups54$ & $-28\arcd29\arcm11\zdot\arcs3$ & 1.806 & 16.978 & 0.351 & 0.203 & 5.214180 \\ 
21 & 7051 & $18\uph00\upm42\zdot\ups73$ & $-28\arcd27\arcm10\zdot\arcs0$ & 1.867 & 17.868 & 0.323 & 0.204 & 2.701986 \\ 
21 & 7356 & $18\uph00\upm25\zdot\ups77$ & $-28\arcd24\arcm37\zdot\arcs8$ & 1.944 & 16.954 & 0.316 & 0.307 & 4.641814 \\ 
22 &   10 & $17\uph56\upm27\zdot\ups07$ & $-31\arcd15\arcm31\zdot\arcs6$ & 1.988 & 17.542 & 0.285 & 0.196 & 4.378102 \\ 
22 &  114 & $17\uph57\upm20\zdot\ups34$ & $-31\arcd14\arcm52\zdot\arcs1$ & 2.215 & 17.198 & 0.392 & 0.203 & 4.809705 \\ 
22 &  222 & $17\uph56\upm23\zdot\ups41$ & $-31\arcd13\arcm10\zdot\arcs8$ & 2.259 & 17.795 & 0.155 & 0.057 & 11.372206 \\ 
22 & 1335 & $17\uph57\upm12\zdot\ups80$ & $-31\arcd01\arcm58\zdot\arcs1$ & 1.774 & 17.769 & 0.405 & 0.211 & 2.119743 \\ 
22 & 2938 & $17\uph56\upm28\zdot\ups02$ & $-30\arcd47\arcm05\zdot\arcs7$ & 1.761 & 16.769 & 0.240 & 0.113 & 4.996538 \\ 
23 &  293 & $17\uph57\upm23\zdot\ups92$ & $-31\arcd36\arcm32\zdot\arcs2$ & 2.103 & 18.642 & 0.208 & 0.083 & 5.482078 \\ 
23 & 1425 & $17\uph57\upm34\zdot\ups18$ & $-31\arcd23\arcm01\zdot\arcs8$ & 2.346 & 17.864 & 0.274 & 0.241 & 2.203385 \\ 
23 & 1695 & $17\uph57\upm38\zdot\ups45$ & $-31\arcd19\arcm27\zdot\arcs1$ & 2.601 & 17.894 & 0.276 & 0.219 & 2.837150 \\ 
23 & 3566 & $17\uph57\upm44\zdot\ups81$ & $-30\arcd58\arcm12\zdot\arcs4$ & 2.299 & 18.165 & 0.446 & 0.227 & 2.023772 \\ 
23 & 3832 & $17\uph57\upm38\zdot\ups48$ & $-30\arcd55\arcm07\zdot\arcs3$ & 2.678 & 15.592 & 0.481 & 0.085 & 36.548702 \\ 
24 & 1797 & $17\uph53\upm38\zdot\ups44$ & $-32\arcd55\arcm12\zdot\arcs2$ & 2.144 & 17.542 & 0.350 & 0.227 & 8.378970 \\ 
24 & 2166 & $17\uph53\upm03\zdot\ups75$ & $-32\arcd49\arcm24\zdot\arcs8$ & 2.029 & 16.710 & 0.304 & 0.147 & 6.788291 \\ 
24 & 2463 & $17\uph53\upm04\zdot\ups50$ & $-32\arcd45\arcm05\zdot\arcs9$ & 2.564 & 15.956 & 0.266 & 0.075 & 2.846186 \\ 
25 &  425 & $17\uph54\upm39\zdot\ups82$ & $-33\arcd11\arcm54\zdot\arcs2$ & 2.022 & 17.896 & 0.357 & 0.308 & 2.807364 \\ 
25 & 2256 & $17\uph54\upm16\zdot\ups80$ & $-32\arcd37\arcm23\zdot\arcs3$ & 1.886 & 17.655 & 0.446 & 0.233 & 3.888740 \\ 
25 & 2800 & $17\uph54\upm22\zdot\ups11$ & $-32\arcd28\arcm52\zdot\arcs9$ & 1.867 & 17.709 & 0.234 & 0.117 & 4.863817 \\ 
25 & 2836 & $17\uph53\upm59\zdot\ups11$ & $-32\arcd28\arcm01\zdot\arcs3$ & 1.870 & 17.235 & 0.357 & 0.255 & 2.113844 \\ 
26 &  572 & $17\uph46\upm59\zdot\ups47$ & $-35\arcd19\arcm42\zdot\arcs8$ & 2.003 & 16.379 & 0.215 & 0.084 & 13.478680 \\ 
26 &  946 & $17\uph46\upm58\zdot\ups93$ & $-35\arcd15\arcm02\zdot\arcs0$ & 1.814 & 17.455 & 0.287 & 0.190 & 2.131286 \\ 
26 & 1376 & $17\uph47\upm03\zdot\ups47$ & $-35\arcd09\arcm54\zdot\arcs2$ & 1.838 & 17.619 & 0.348 & 0.312 & 2.009605 \\ 
26 & 3703 & $17\uph47\upm10\zdot\ups75$ & $-34\arcd43\arcm40\zdot\arcs4$ & 1.682 & 15.740 & 0.116 & 0.029 & 2.620312 \\ 
26 & 3875 & $17\uph47\upm08\zdot\ups15$ & $-34\arcd43\arcm16\zdot\arcs5$ & 1.732 & 17.322 & 0.117 & 0.076 & 2.396786 \\ 
26 & 4375 & $17\uph47\upm45\zdot\ups91$ & $-34\arcd36\arcm39\zdot\arcs6$ & 2.014 & 18.062 & 0.197 & 0.122 & 2.961308 \\ 
26 & 4395 & $17\uph46\upm46\zdot\ups31$ & $-34\arcd36\arcm26\zdot\arcs1$ & 2.021 & 17.902 & 0.171 & 0.125 & 3.688354 \\ 
27 &  323 & $17\uph48\upm23\zdot\ups02$ & $-35\arcd32\arcm00\zdot\arcs8$ & 1.524 & 16.950 & 0.120 & 0.051 & 2.038052 \\ 
27 &  646 & $17\uph48\upm20\zdot\ups08$ & $-35\arcd26\arcm48\zdot\arcs3$ & 1.599 & 17.457 & 0.357 & 0.325 & 3.404542 \\ 
27 & 1512 & $17\uph48\upm30\zdot\ups63$ & $-35\arcd15\arcm20\zdot\arcs6$ & 1.594 & 16.877 & 0.312 & 0.284 & 3.744963 \\ 
27 & 1587 & $17\uph47\upm58\zdot\ups47$ & $-35\arcd13\arcm20\zdot\arcs4$ & 1.638 & 17.604 & 0.290 & 0.132 & 2.221579 \\ 
27 & 2801 & $17\uph47\upm57\zdot\ups32$ & $-34\arcd55\arcm17\zdot\arcs3$ & 1.639 & 17.837 & 0.257 & 0.127 & 4.287232 \\ 
28 &  974 & $17\uph46\upm33\zdot\ups81$ & $-36\arcd56\arcm26\zdot\arcs9$ & 1.694 & 17.141 & 0.181 & 0.059 & 3.625968 \\ 
28 & 1355 & $17\uph46\upm59\zdot\ups43$ & $-36\arcd43\arcm51\zdot\arcs2$ & 1.735 & 18.333 & 0.271 & 0.222 & 2.050652 \\ 
29 &  491 & $17\uph47\upm48\zdot\ups95$ & $-37\arcd21\arcm59\zdot\arcs4$ & 1.596 & 16.120 & 0.386 & 0.212 & 3.137304 \\ 
29 &  943 & $17\uph47\upm53\zdot\ups46$ & $-37\arcd11\arcm14\zdot\arcs8$ & 1.603 & 17.413 & 0.226 & 0.125 & 6.781612 \\ 
29 & 1001 & $17\uph48\upm13\zdot\ups64$ & $-37\arcd10\arcm05\zdot\arcs6$ & 1.461 & 16.225 & 0.303 & 0.238 & 3.450851 \\ 
29 & 1063 & $17\uph47\upm57\zdot\ups24$ & $-37\arcd08\arcm22\zdot\arcs1$ & 1.574 & 16.961 & 0.398 & 0.165 & 2.255049 \\ 
29 & 1149 & $17\uph48\upm43\zdot\ups11$ & $-37\arcd06\arcm33\zdot\arcs9$ & 1.812 & 17.158 & 0.220 & 0.207 & 8.651742 \\ 
29 & 1755 & $17\uph48\upm12\zdot\ups65$ & $-36\arcd52\arcm44\zdot\arcs1$ & 1.949 & 16.930 & 0.463 & 0.208 & 10.219748 \\ 
30 &  983 & $18\uph01\upm12\zdot\ups34$ & $-29\arcd08\arcm57\zdot\arcs4$ & 1.968 & 16.868 & 0.101 & 0.075 & 36.279276 \\ 
30 & 1034 & $18\uph01\upm51\zdot\ups72$ & $-29\arcd08\arcm47\zdot\arcs0$ & 1.971 & 17.219 & 0.304 & 0.153 & 3.515654 \\ 
30 & 1558 & $18\uph00\upm57\zdot\ups01$ & $-29\arcd03\arcm37\zdot\arcs4$ & 1.763 & 16.068 & 0.139 & 0.043 & 5.385688 \\ 
\hline}
\setcounter{table}{0}
\MakeTableSepp{rrccccccr}{12.5cm}{Continued}
{\hline
\douprule
Field & ID & RA & DEC & $(V-I)$ & $I$ & $R_1/a$ & $R_2/a$ & period [days]\\
\hline
30 & 2126 & $18\uph01\upm53\zdot\ups58$ & $-28\arcd59\arcm19\zdot\arcs6$ & 1.889 & 17.735 & 0.396 & 0.313 & 2.539593 \\ 
30 & 2158 & $18\uph01\upm07\zdot\ups93$ & $-28\arcd58\arcm25\zdot\arcs6$ & 1.642 & 16.387 & 0.390 & 0.293 & 2.205462 \\ 
30 & 2508 & $18\uph01\upm24\zdot\ups79$ & $-28\arcd55\arcm21\zdot\arcs9$ & 1.791 & 17.463 & 0.311 & 0.157 & 5.518564 \\ 
30 & 3167 & $18\uph01\upm48\zdot\ups86$ & $-28\arcd51\arcm10\zdot\arcs8$ & 1.615 & 17.629 & 0.160 & 0.111 & 3.254480 \\ 
30 & 5566 & $18\uph01\upm11\zdot\ups19$ & $-28\arcd32\arcm38\zdot\arcs6$ & 1.661 & 17.393 & 0.193 & 0.134 & 4.206418 \\ 
30 & 5978 & $18\uph00\upm57\zdot\ups98$ & $-28\arcd29\arcm09\zdot\arcs2$ & 1.774 & 16.501 & 0.173 & 0.067 & 10.134768 \\ 
31 &  399 & $18\uph01\upm53\zdot\ups58$ & $-28\arcd59\arcm19\zdot\arcs7$ & 1.831 & 17.735 & 0.389 & 0.314 & 2.539666 \\ 
31 & 2277 & $18\uph02\upm26\zdot\ups01$ & $-28\arcd34\arcm28\zdot\arcs9$ & 2.109 & 17.576 & 0.311 & 0.179 & 3.181014 \\ 
31 & 2670 & $18\uph02\upm30\zdot\ups81$ & $-28\arcd29\arcm58\zdot\arcs1$ & 1.951 & 17.559 & 0.412 & 0.260 & 2.312017 \\ 
31 & 2793 & $18\uph01\upm53\zdot\ups97$ & $-28\arcd27\arcm47\zdot\arcs4$ & 1.789 & 17.137 & 0.201 & 0.105 & 6.405328 \\ 
31 & 3308 & $18\uph02\upm27\zdot\ups26$ & $-28\arcd23\arcm22\zdot\arcs3$ & 1.777 & 17.224 & 0.365 & 0.180 & 2.686944 \\ 
31 & 4092 & $18\uph02\upm49\zdot\ups03$ & $-28\arcd16\arcm33\zdot\arcs3$ & 1.659 & 17.698 & 0.295 & 0.231 & 3.067622 \\ 
31 & 4328 & $18\uph02\upm25\zdot\ups28$ & $-28\arcd13\arcm40\zdot\arcs1$ & 1.593 & 17.387 & 0.104 & 0.070 & 7.872418 \\ 
31 & 4787 & $18\uph02\upm49\zdot\ups76$ & $-28\arcd09\arcm59\zdot\arcs3$ & 1.556 & 17.561 & 0.420 & 0.304 & 3.079226 \\ 
32 &  926 & $18\uph03\upm26\zdot\ups66$ & $-28\arcd54\arcm40\zdot\arcs6$ & 1.544 & 17.382 & 0.107 & 0.053 & 13.887762 \\ 
32 & 1504 & $18\uph03\upm37\zdot\ups95$ & $-28\arcd48\arcm24\zdot\arcs6$ & 1.680 & 17.251 & 0.273 & 0.167 & 6.526399 \\ 
32 & 1928 & $18\uph03\upm54\zdot\ups07$ & $-28\arcd43\arcm54\zdot\arcs9$ & 1.738 & 16.620 & 0.383 & 0.251 & 4.992396 \\ 
32 & 2053 & $18\uph03\upm34\zdot\ups10$ & $-28\arcd41\arcm56\zdot\arcs9$ & 1.686 & 17.126 & 0.172 & 0.111 & 7.414870 \\ 
32 & 3061 & $18\uph03\upm17\zdot\ups72$ & $-28\arcd31\arcm15\zdot\arcs5$ & 2.076 & 17.038 & 0.266 & 0.110 & 5.346271 \\ 
32 & 3753 & $18\uph03\upm19\zdot\ups86$ & $-28\arcd23\arcm36\zdot\arcs5$ & 1.752 & 17.151 & 0.184 & 0.054 & 9.886980 \\ 
32 & 4589 & $18\uph03\upm56\zdot\ups25$ & $-28\arcd14\arcm43\zdot\arcs9$ & 1.569 & 16.890 & 0.153 & 0.087 & 5.627108 \\ 
33 &  547 & $18\uph05\upm36\zdot\ups60$ & $-29\arcd13\arcm14\zdot\arcs3$ & 1.840 & 17.417 & 0.383 & 0.170 & 3.463207 \\ 
33 &  686 & $18\uph05\upm26\zdot\ups29$ & $-29\arcd11\arcm57\zdot\arcs4$ & 1.612 & 17.791 & 0.162 & 0.097 & 4.095290 \\ 
33 & 1629 & $18\uph05\upm21\zdot\ups08$ & $-28\arcd59\arcm58\zdot\arcs8$ & 1.760 & 17.658 & 0.305 & 0.162 & 2.058984 \\ 
33 & 2278 & $18\uph05\upm26\zdot\ups10$ & $-28\arcd52\arcm17\zdot\arcs1$ & 1.663 & 18.171 & 0.396 & 0.165 & 2.013036 \\ 
33 & 3188 & $18\uph05\upm45\zdot\ups64$ & $-28\arcd42\arcm47\zdot\arcs6$ & 1.867 & 17.858 & 0.265 & 0.191 & 6.148376 \\ 
34 &  420 & $17\uph58\upm18\zdot\ups98$ & $-29\arcd32\arcm57\zdot\arcs1$ & 2.097 & 17.302 & 0.277 & 0.139 & 2.231450 \\ 
34 &  793 & $17\uph57\upm50\zdot\ups07$ & $-29\arcd29\arcm10\zdot\arcs9$ & 1.857 & 17.068 & 0.310 & 0.200 & 7.976383 \\ 
34 & 1615 & $17\uph58\upm36\zdot\ups95$ & $-29\arcd22\arcm51\zdot\arcs7$ & 2.005 & 17.930 & 0.376 & 0.265 & 2.540265 \\ 
34 & 1706 & $17\uph58\upm20\zdot\ups31$ & $-29\arcd22\arcm39\zdot\arcs9$ & 1.919 & 17.476 & 0.283 & 0.079 & 4.267768 \\ 
34 & 1951 & $17\uph58\upm35\zdot\ups51$ & $-29\arcd20\arcm59\zdot\arcs3$ & 1.838 & 17.803 & 0.313 & 0.295 & 2.224765 \\ 
34 & 2456 & $17\uph58\upm05\zdot\ups20$ & $-29\arcd16\arcm05\zdot\arcs0$ & 2.114 & 17.097 & 0.248 & 0.080 & 8.471958 \\ 
34 & 2580 & $17\uph58\upm20\zdot\ups21$ & $-29\arcd14\arcm57\zdot\arcs9$ & 1.779 & 17.759 & 0.127 & 0.093 & 6.439528 \\ 
34 & 2823 & $17\uph58\upm33\zdot\ups05$ & $-29\arcd13\arcm31\zdot\arcs2$ & 1.583 & 16.233 & 0.160 & 0.045 & 3.521810 \\ 
34 & 2973 & $17\uph58\upm42\zdot\ups81$ & $-29\arcd12\arcm15\zdot\arcs3$ & 1.649 & 16.020 & 0.369 & 0.202 & 3.183053 \\ 
34 & 5744 & $17\uph58\upm04\zdot\ups66$ & $-28\arcd53\arcm04\zdot\arcs9$ & 2.026 & 17.878 & 0.158 & 0.089 & 5.301156 \\ 
35 & 2440 & $18\uph04\upm06\zdot\ups62$ & $-27\arcd58\arcm54\zdot\arcs6$ & 1.793 & 17.764 & 0.312 & 0.198 & 2.022147 \\ 
35 & 2508 & $18\uph04\upm02\zdot\ups60$ & $-27\arcd58\arcm23\zdot\arcs6$ & 1.722 & 14.118 & 0.210 & 0.031 & 21.505206 \\ 
35 & 2897 & $18\uph04\upm36\zdot\ups18$ & $-27\arcd54\arcm27\zdot\arcs0$ & 2.493 & 16.517 & 0.335 & 0.222 & 2.453367 \\ 
35 & 4121 & $18\uph04\upm27\zdot\ups93$ & $-27\arcd41\arcm05\zdot\arcs1$ & 1.864 & 16.660 & 0.437 & 0.248 & 2.551206 \\ 
35 & 4556 & $18\uph04\upm38\zdot\ups36$ & $-27\arcd36\arcm26\zdot\arcs2$ & 1.624 & 17.057 & 0.200 & 0.092 & 4.054498 \\ 
35 & 4721 & $18\uph04\upm11\zdot\ups40$ & $-27\arcd33\arcm39\zdot\arcs3$ & 1.841 & 15.595 & 0.317 & 0.121 & 2.759243 \\ 
36 & 1929 & $18\uph05\upm47\zdot\ups36$ & $-28\arcd13\arcm03\zdot\arcs3$ & 1.910 & 16.848 & 0.393 & 0.173 & 8.543916 \\ 
36 & 3762 & $18\uph05\upm05\zdot\ups62$ & $-28\arcd00\arcm35\zdot\arcs3$ & 1.639 & 16.801 & 0.128 & 0.050 & 2.586147 \\ 
36 & 4333 & $18\uph05\upm31\zdot\ups34$ & $-27\arcd57\arcm11\zdot\arcs9$ & 1.964 & 17.726 & 0.389 & 0.138 & 3.928802 \\ 
36 & 4493 & $18\uph05\upm51\zdot\ups25$ & $-27\arcd56\arcm20\zdot\arcs9$ & 1.701 & 17.319 & 0.292 & 0.249 & 3.118629 \\ 
36 & 4494 & $18\uph05\upm55\zdot\ups08$ & $-27\arcd56\arcm12\zdot\arcs2$ & 1.992 & 17.223 & 0.330 & 0.271 & 2.910082 \\ 
\hline}
\renewcommand{\arraystretch}{0.85}
\setcounter{table}{0}
\MakeTableSepp{rrccccccr}{12.5cm}{Concluded}
{\hline
\douprule
Field & ID & RA & DEC & $(V-I)$ & $I$ & $R_1/a$ & $R_2/a$ & period [days]\\
\hline
36 & 7720 & $18\uph05\upm37\zdot\ups14$ & $-27\arcd36\arcm22\zdot\arcs0$ & 1.613 & 17.819 & 0.239 & 0.161 & 3.115270 \\ 
37 &  956 & $17\uph52\upm52\zdot\ups82$ & $-30\arcd18\arcm52\zdot\arcs1$ & 2.442 & 17.944 & 0.334 & 0.254 & 3.262918 \\ 
37 & 2068 & $17\uph52\upm22\zdot\ups28$ & $-30\arcd09\arcm18\zdot\arcs0$ & 2.722 & 18.234 & 0.388 & 0.296 & 3.662674 \\ 
37 & 2768 & $17\uph52\upm19\zdot\ups88$ & $-30\arcd05\arcm22\zdot\arcs0$ & 2.730 & 17.333 & 0.391 & 0.318 & 5.733310 \\ 
37 & 3186 & $17\uph52\upm35\zdot\ups96$ & $-30\arcd02\arcm52\zdot\arcs3$ & 3.339 & 17.570 & 0.310 & 0.211 & 5.098516 \\ 
37 & 3240 & $17\uph52\upm05\zdot\ups16$ & $-30\arcd01\arcm32\zdot\arcs7$ & 3.122 & 18.193 & 0.250 & 0.199 & 3.091703 \\ 
37 & 4541 & $17\uph52\upm06\zdot\ups26$ & $-29\arcd53\arcm15\zdot\arcs8$ & 2.800 & 16.147 & 0.435 & 0.076 & 2.907292 \\ 
37 & 6333 & $17\uph52\upm13\zdot\ups81$ & $-29\arcd40\arcm19\zdot\arcs7$ & 2.973 & 18.570 & 0.290 & 0.267 & 2.943085 \\ 
37 & 6549 & $17\uph52\upm52\zdot\ups89$ & $-29\arcd39\arcm27\zdot\arcs2$ & 2.724 & 17.381 & 0.275 & 0.164 & 2.647908 \\ 
37 & 7240 & $17\uph52\upm02\zdot\ups98$ & $-29\arcd35\arcm30\zdot\arcs7$ & 3.491 & 17.082 & 0.300 & 0.087 & 2.012917 \\ 
37 & 8237 & $17\uph52\upm51\zdot\ups31$ & $-29\arcd30\arcm29\zdot\arcs8$ & 2.834 & 17.937 & 0.426 & 0.259 & 3.955137 \\ 
38 & 1003 & $18\uph01\upm29\zdot\ups06$ & $-30\arcd14\arcm16\zdot\arcs6$ & 1.901 & 17.678 & 0.321 & 0.236 & 2.495866 \\ 
38 & 1675 & $18\uph01\upm37\zdot\ups77$ & $-30\arcd07\arcm31\zdot\arcs4$ & 1.619 & 17.156 & 0.343 & 0.296 & 2.396601 \\ 
38 & 3532 & $18\uph01\upm10\zdot\ups10$ & $-29\arcd46\arcm17\zdot\arcs9$ & 1.795 & 17.420 & 0.357 & 0.207 & 6.338837 \\ 
38 & 3758 & $18\uph02\upm00\zdot\ups41$ & $-29\arcd44\arcm30\zdot\arcs0$ & 1.634 & 17.102 & 0.234 & 0.076 & 4.501576 \\ 
38 & 4718 & $18\uph01\upm24\zdot\ups01$ & $-29\arcd33\arcm20\zdot\arcs9$ & 1.868 & 16.910 & 0.330 & 0.106 & 4.589060 \\ 
38 & 5059 & $18\uph01\upm57\zdot\ups05$ & $-29\arcd30\arcm27\zdot\arcs5$ & 2.193 & 16.816 & 0.108 & 0.105 & 3.617376 \\ 
39 &  511 & $17\uph56\upm01\zdot\ups79$ & $-30\arcd08\arcm43\zdot\arcs3$ & 2.124 & 17.210 & 0.323 & 0.143 & 2.220596 \\ 
39 & 1893 & $17\uph55\upm08\zdot\ups38$ & $-29\arcd56\arcm48\zdot\arcs4$ & 1.872 & 17.722 & 0.396 & 0.146 & 2.452596 \\ 
39 & 2555 & $17\uph55\upm24\zdot\ups39$ & $-29\arcd52\arcm20\zdot\arcs7$ & 2.389 & 17.157 & 0.288 & 0.200 & 2.879248 \\ 
39 & 2745 & $17\uph55\upm16\zdot\ups18$ & $-29\arcd50\arcm20\zdot\arcs9$ & 2.168 & 16.717 & 0.313 & 0.088 & 8.139266 \\ 
39 & 3008 & $17\uph55\upm31\zdot\ups48$ & $-29\arcd48\arcm50\zdot\arcs6$ & 2.772 & 18.302 & 0.268 & 0.154 & 3.591320 \\ 
39 & 4483 & $17\uph55\upm25\zdot\ups33$ & $-29\arcd38\arcm32\zdot\arcs6$ & 2.412 & 18.056 & 0.245 & 0.165 & 4.946640 \\ 
39 & 5279 & $17\uph55\upm15\zdot\ups81$ & $-29\arcd32\arcm08\zdot\arcs1$ & 1.833 & 16.698 & 0.306 & 0.118 & 5.776876 \\ 
39 & 5315 & $17\uph55\upm37\zdot\ups41$ & $-29\arcd31\arcm44\zdot\arcs2$ & 1.905 & 18.374 & 0.292 & 0.180 & 2.048580 \\ 
39 & 6095 & $17\uph55\upm19\zdot\ups90$ & $-29\arcd26\arcm12\zdot\arcs8$ & 2.106 & 17.206 & 0.388 & 0.236 & 2.885993 \\ 
39 & 6554 & $17\uph56\upm05\zdot\ups16$ & $-29\arcd23\arcm08\zdot\arcs8$ & 2.351 & 18.098 & 0.312 & 0.228 & 2.144490 \\ 
40 & 1732 & $17\uph51\upm09\zdot\ups87$ & $-33\arcd19\arcm16\zdot\arcs9$ & 2.229 & 16.557 & 0.351 & 0.250 & 2.934772 \\ 
40 & 1808 & $17\uph50\upm54\zdot\ups17$ & $-33\arcd18\arcm35\zdot\arcs6$ & 2.199 & 17.029 & 0.370 & 0.259 & 2.465588 \\ 
41 &  207 & $17\uph52\upm06\zdot\ups31$ & $-33\arcd31\arcm42\zdot\arcs0$ & 2.280 & 16.931 & 0.440 & 0.241 & 9.307920 \\ 
41 &  279 & $17\uph52\upm37\zdot\ups85$ & $-33\arcd30\arcm41\zdot\arcs5$ & 1.833 & 18.366 & 0.196 & 0.139 & 2.583558 \\ 
41 & 1414 & $17\uph52\upm15\zdot\ups20$ & $-33\arcd14\arcm55\zdot\arcs5$ & 2.076 & 17.974 & 0.312 & 0.148 & 2.544096 \\ 
41 & 1593 & $17\uph52\upm35\zdot\ups87$ & $-33\arcd13\arcm14\zdot\arcs6$ & 2.074 & 17.135 & 0.409 & 0.225 & 6.934563 \\ 
42 &  872 & $18\uph09\upm09\zdot\ups69$ & $-27\arcd09\arcm24\zdot\arcs1$ & 2.014 & 17.096 & 0.242 & 0.120 & 2.083840 \\ 
42 &  875 & $18\uph09\upm11\zdot\ups46$ & $-27\arcd09\arcm32\zdot\arcs6$ & 1.852 & 17.414 & 0.239 & 0.110 & 4.297795 \\ 
42 & 1309 & $18\uph09\upm08\zdot\ups77$ & $-27\arcd03\arcm57\zdot\arcs7$ & 1.706 & 16.405 & 0.446 & 0.261 & 4.606988 \\ 
42 & 4197 & $18\uph09\upm17\zdot\ups51$ & $-26\arcd26\arcm31\zdot\arcs4$ & 2.056 & 17.903 & 0.416 & 0.158 & 2.161317 \\ 
42 & 4324 & $18\uph08\upm55\zdot\ups48$ & $-26\arcd25\arcm19\zdot\arcs7$ & 1.976 & 15.675 & 0.373 & 0.224 & 2.452289 \\ 
43 &  305 & $17\uph34\upm47\zdot\ups64$ & $-27\arcd32\arcm53\zdot\arcs8$ & 2.926 & 18.133 & 0.267 & 0.174 & 6.457909 \\ 
43 &  310 & $17\uph35\upm04\zdot\ups55$ & $-27\arcd32\arcm29\zdot\arcs1$ & 2.706 & 18.111 & 0.348 & 0.223 & 6.223798 \\ 
43 &  834 & $17\uph34\upm45\zdot\ups85$ & $-27\arcd22\arcm45\zdot\arcs7$ & 2.998 & 18.724 & 0.349 & 0.309 & 2.242491 \\ 
43 & 1023 & $17\uph35\upm38\zdot\ups97$ & $-27\arcd19\arcm56\zdot\arcs0$ & 2.769 & 18.727 & 0.410 & 0.216 & 2.140171 \\ 
43 & 2852 & $17\uph35\upm40\zdot\ups60$ & $-26\arcd53\arcm12\zdot\arcs4$ & 2.654 & 17.204 & 0.306 & 0.263 & 4.098304 \\ 
44 & 2264 & $17\uph49\upm11\zdot\ups53$ & $-30\arcd10\arcm08\zdot\arcs3$ & 3.519 & 18.243 & 0.152 & 0.117 & 3.218736 \\ 
46 &  178 & $18\uph04\upm24\zdot\ups67$ & $-30\arcd28\arcm10\zdot\arcs5$ & 1.874 & 16.991 & 0.311 & 0.156 & 3.379805 \\ 
46 & 1863 & $18\uph04\upm26\zdot\ups40$ & $-29\arcd42\arcm49\zdot\arcs6$ & 1.402 & 16.447 & 0.137 & 0.077 & 5.429178 \\ 
46 & 1982 & $18\uph04\upm46\zdot\ups68$ & $-29\arcd39\arcm41\zdot\arcs2$ & 1.796 & 16.902 & 0.371 & 0.227 & 3.712870 \\ 
46 & 2039 & $18\uph04\upm16\zdot\ups65$ & $-29\arcd37\arcm23\zdot\arcs5$ & 2.096 & 15.451 & 0.401 & 0.201 & 2.679111 \\ 
48 &  911 & $17\uph28\upm30\zdot\ups26$ & $-39\arcd21\arcm47\zdot\arcs3$ & 1.924 & 16.172 & 0.376 & 0.264 & 3.618835 \\ 
\hline}

\renewcommand{\arraystretch}{1.0}
\setcounter{table}{1}
\MakeTableSepp{c@{\hspace{3pt}}r@{\hspace{7pt}}c@{\hspace{7pt}}c@{\hspace{7pt}}c@{\hspace{7pt}}c
c@{\hspace{5pt}}
c@{\hspace{3pt}}r@{\hspace{7pt}}c@{\hspace{7pt}}c@{\hspace{7pt}}c@{\hspace{7pt}}c}{12.5cm}{Parameters
for the gold (1$^{\rm st}$ part) and silver (2$^{\rm nd}$ part) of the identified RG
eclipsing binary pairs}
{\cline{1-6}  \cline{8-13}
\noalign{\vskip3pt}
Field& \multicolumn{1}{c}{ID} & $e$ & $E_2$ & $t_1$ [d] & $\sin i$ &&Field& \multicolumn{1}{c}{ID} & $e$ & $E_2$ & $t_1$ [d] & $\sin i$ \\
\noalign{\vskip3pt}
\cline{1-6}  \cline{8-13}
\noalign{\vskip3pt}
3  & 2325 & 0.146 & 2450000.18 & 0.53 & 0.934 &&  3  & 1149 & 0.043 & 2450002.60 & 0.43 & 0.932 \\ 
3  & 2823 & 0.016 & 2449999.60 & 0.41 & 0.996 &&  3  & 1688 & 0.113 & 2450002.78 & 0.30 & 0.993 \\ 
3  & 2929 & 0.007 & 2450002.04 & 0.41 & 0.982 &&  3  & 2195 & 0.113 & 2449999.43 & 0.41 & 0.912 \\ 
3  & 7715 & 0.085 & 2450000.38 & 0.64 & 0.978 &&  3  & 3488 & 0.141 & 2450000.73 & 0.18 & 1.000 \\ 
4  &  830 & 0.062 & 2450002.58 & 0.91 & 0.915 &&  3  & 3489 & 0.161 & 2450000.85 & 0.27 & 0.999 \\ 
4  &  831 & 0.071 & 2450000.26 & 0.30 & 0.999 &&  3  & 3547 & 0.143 & 2450000.24 & 0.21 & 0.988 \\ 
4  & 1290 & 0.105 & 2450003.48 & 0.45 & 0.995 &&  3  & 3744 & 0.103 & 2450001.63 & 0.37 & 0.964 \\ 
5  & 1889 & 0.133 & 2449998.25 & 0.81 & 0.985 &&  3  & 6413 & 0.169 & 2450005.48 & 0.87 & 0.978 \\ 
7  & 1089 & 0.049 & 2450001.36 & 0.31 & 1.000 &&  3  & 8222 & 0.212 & 2450006.10 & 0.72 & 0.983 \\ 
9  &  132 & 0.150 & 2450001.36 & 0.43 & 0.947 &&  3  & 8395 & 0.088 & 2449999.98 & 0.71 & 0.990 \\ 
9  & 1502 & 0.008 & 2450002.67 & 0.42 & 0.984 &&  4  & 1047 & 0.061 & 2450001.89 & 0.21 & 0.975 \\ 
13 & 1282 & 0.036 & 2450000.63 & 0.99 & 0.990 &&  4  & 1120 & 0.326 & 2450004.47 & 1.02 & 0.941 \\ 
14 & 3912 & 0.189 & 2450001.04 & 0.61 & 0.974 &&  4  & 1224 & 0.340 & 2450000.85 & 0.61 & 0.987 \\ 
24 & 1374 & 0.310 & 2450000.38 & 0.61 & 0.943 &&  4  & 1289 & 0.006 & 2450000.10 & 0.30 & 0.991 \\ 
24 & 2461 & 0.009 & 2449990.80 & 2.25 & 0.955 &&  4  & 3707 & 0.010 & 2450002.88 & 0.23 & 0.983 \\ 
30 & 1600 & 0.009 & 2450002.26 & 0.41 & 0.977 &&  4  & 4049 & 0.056 & 2450000.30 & 0.15 & 0.995 \\ 
30 & 6658 & 0.061 & 2450007.81 & 1.11 & 0.960 &&  4  & 4449 & 0.283 & 2450002.19 & 0.39 & 0.925 \\ 
31 &  631 & 0.270 & 2450002.81 & 0.46 & 0.981 &&  4  & 4564 & 0.098 & 2450001.56 & 0.35 & 0.987 \\ 
31 & 2641 & 0.022 & 2450002.04 & 0.49 & 0.992 &&  4  & 5392 & 0.043 & 2450001.82 & 0.15 & 0.991 \\ 
34 & 1369 & 0.030 & 2450000.31 & 0.79 & 0.988 &&  4  & 6839 & 0.004 & 2450000.51 & 0.24 & 0.961 \\ 
34 & 6500 & 0.091 & 2450002.73 & 0.39 & 0.996 &&  4  & 6905 & 0.028 & 2450001.74 & 0.34 & 0.981 \\ 
37 & 1846 & 0.048 & 2450000.90 & 0.32 & 0.941 &&  4  & 7346 & 0.128 & 2450002.42 & 0.28 & 0.996 \\ 
37 & 6396 & 0.077 & 2450002.06 & 0.72 & 0.967 &&  4  & 8604 & 0.003 & 2449999.09 & 0.27 & 0.998 \\ 
37 & 7093 & 0.057 & 2450002.88 & 0.55 & 0.985 &&  5  & 5124 & 0.074 & 2450001.96 & 0.30 & 0.995 \\ 
39 & 1159 & 0.160 & 2450002.16 & 0.52 & 0.986 &&  6  &  358 & 0.585 & 2450001.97 & 0.27 & 0.993 \\ 
39 & 1604 & 0.084 & 2450000.11 & 0.32 & 1.000 &&  6  &  598 & 0.063 & 2450000.40 & 0.25 & 0.995 \\ 
40 &  310 & 0.293 & 2450003.46 & 0.96 & 0.994 &&  6  & 1214 & 0.191 & 2450003.06 & 0.77 & 0.951 \\ 
40 &  621 & 0.149 & 2450006.12 & 0.96 & 0.953 &&  6  & 1517 & 0.005 & 2450001.71 & 0.21 & 1.000 \\ 
43 & 3095 & 0.015 & 2450001.83 & 0.54 & 0.957 &&  7  &  299 & 0.037 & 2449999.88 & 0.11 & 1.000 \\ 
45 & 1064 & 0.203 & 2450087.27 & 4.99 & 0.999 &&  7  &  605 & 0.134 & 2450002.56 & 0.52 & 0.998 \\ 
45 & 1155 & 0.008 & 2450004.72 & 0.55 & 0.997 &&  7  & 1494 & 0.036 & 2450001.95 & 0.14 & 0.997 \\ 
45 & 1522 & 0.054 & 2450002.05 & 0.28 & 0.993 &&  8  &  430 & 0.149 & 2450001.26 & 0.46 & 0.969 \\ 
45 & 2170 & 0.088 & 2450009.51 & 1.86 & 1.000 &&  8  &  829 & 0.025 & 2450000.33 & 0.31 & 0.995 \\ 
47 &  481 & 0.091 & 2450001.98 & 0.38 & 0.984 &&  10 & 1924 & 0.095 & 2450000.11 & 0.28 & 1.000 \\ 
\cline{1-6}
1  & 1713 & 0.017 & 2450003.65 & 0.25 & 0.998 &&  10 & 2053 & 0.076 & 2450002.24 & 0.37 & 0.990 \\ 
1  & 2513 & 0.003 & 2450002.41 & 0.31 & 0.896 &&  11 & 1231 & 0.017 & 2450004.38 & 0.41 & 0.995 \\ 
1  & 2938 & 0.205 & 2449999.27 & 0.34 & 0.987 &&  11 & 1934 & 0.077 & 2450004.78 & 1.06 & 0.986 \\ 
1  & 3738 & 0.088 & 2450000.55 & 0.30 & 0.995 &&  12 & 1664 & 0.007 & 2450001.58 & 0.16 & 0.999 \\ 
1  & 3846 & 0.377 & 2450002.09 & 0.71 & 0.999 &&  12 & 2249 & 0.029 & 2450002.26 & 0.23 & 1.000 \\ 
1  & 3859 & 0.131 & 2450000.92 & 0.36 & 0.972 &&  12 & 3118 & 0.054 & 2450002.24 & 0.19 & 1.000 \\ 
2  &  892 & 0.284 & 2450001.93 & 0.59 & 0.965 &&  13 &  515 & 0.494 & 2450000.50 & 0.15 & 0.991 \\ 
2  & 1301 & 0.059 & 2450006.32 & 0.52 & 0.995 &&  13 &  643 & 0.242 & 2450001.77 & 0.35 & 0.994 \\ 
2  & 1800 & 0.161 & 2449999.68 & 0.22 & 0.995 &&  13 &  748 & 0.005 & 2450000.69 & 0.18 & 0.996 \\ 
2  & 2542 & 0.024 & 2450001.47 & 0.35 & 0.973 &&  14 &  748 & 0.187 & 2450000.90 & 0.28 & 1.000 \\ 
2  & 3673 & 0.450 & 2450012.14 & 1.33 & 1.000 &&  14 &  842 & 0.097 & 2450006.78 & 0.42 & 0.981 \\ 
2  & 3894 & 0.034 & 2450003.90 & 0.36 & 0.992 &&  14 & 1246 & 0.533 & 2450009.26 & 0.86 & 1.000 \\ 
2  & 4754 & 0.045 & 2449999.62 & 0.22 & 0.995 &&  14 & 1360 & 0.159 & 2450002.40 & 0.47 & 1.000 \\ 
\noalign{\vskip3pt}
\cline{1-6}  \cline{8-13}}

\renewcommand{\arraystretch}{1.0}
\setcounter{table}{1}
\MakeTableSepp{c@{\hspace{3pt}}r@{\hspace{7pt}}c@{\hspace{7pt}}c@{\hspace{7pt}}c@{\hspace{7pt}}c c@{\hspace{5pt}} c@{\hspace{3pt}}r@{\hspace{7pt}}c@{\hspace{7pt}}c@{\hspace{7pt}}c@{\hspace{7pt}}c}{12.5cm}{Continued}
{\cline{1-6}  \cline{8-13}
\noalign{\vskip3pt}
Field& \multicolumn{1}{c}{ID} & $e$ & $E_2$ & $t_1$ [d] & $\sin i$ &&Field& \multicolumn{1}{c}{ID} & $e$ & $E_2$ & $t_1$ [d] & $\sin i$ \\
\noalign{\vskip3pt}
\cline{1-6}  \cline{8-13}
\noalign{\vskip3pt}
14 & 2296 & 0.075 & 2450002.07 & 0.19 & 0.995 && 21 & 4732 & 0.715 & 2450001.35 & 0.20 & 0.995 \\ 
14 & 3223 & 0.031 & 2450001.18 & 0.26 & 0.991 && 21 & 6233 & 0.021 & 2450001.27 & 0.39 & 0.998 \\ 
14 & 3755 & 0.165 & 2450001.69 & 0.21 & 0.971 && 21 & 6795 & 0.052 & 2450003.61 & 0.58 & 0.998 \\ 
15 &  529 & 0.270 & 2449999.35 & 0.49 & 0.983 && 21 & 7051 & 0.070 & 2450000.94 & 0.28 & 0.999 \\ 
15 &  631 & 0.027 & 2450000.15 & 0.26 & 1.000 && 21 & 7356 & 0.031 & 2450000.90 & 0.46 & 0.964 \\ 
15 &  851 & 0.000 & 2450001.61 & 0.40 & 0.994 && 22 &   10 & 0.160 & 2450000.71 & 0.40 & 0.979 \\ 
15 & 1341 & 0.014 & 2450001.42 & 0.17 & 0.995 && 22 &  114 & 0.238 & 2450000.75 & 0.60 & 0.988 \\ 
15 & 2256 & 0.212 & 2450002.68 & 0.31 & 0.997 && 22 &  222 & 0.046 & 2450007.33 & 0.56 & 1.000 \\ 
16 &  147 & 0.006 & 2450003.99 & 0.46 & 0.996 && 22 & 1335 & 0.020 & 2450001.18 & 0.27 & 0.958 \\ 
16 &  443 & 0.099 & 2449999.72 & 0.27 & 0.993 && 22 & 2938 & 0.082 & 2450002.79 & 0.38 & 0.992 \\ 
16 &  722 & 0.010 & 2450002.17 & 0.64 & 0.946 && 23 &  293 & 0.118 & 2450004.27 & 0.36 & 0.999 \\ 
16 & 2304 & 0.081 & 2450002.34 & 0.40 & 0.981 && 23 & 1425 & 0.039 & 2449999.64 & 0.19 & 0.987 \\ 
16 & 3066 & 0.180 & 2450002.01 & 0.39 & 0.998 && 23 & 1695 & 0.018 & 2449998.85 & 0.25 & 0.997 \\ 
17 &  417 & 0.047 & 2450000.36 & 0.35 & 0.981 && 23 & 3566 & 0.130 & 2450000.91 & 0.29 & 0.984 \\ 
17 & 1576 & 0.446 & 2450000.33 & 0.46 & 1.000 && 23 & 3832 & 0.042 & 2450036.01 & 5.55 & 0.983 \\ 
17 & 1577 & 0.023 & 2450013.98 & 1.18 & 0.999 && 24 & 1797 & 0.422 & 2450007.07 & 0.93 & 0.985 \\ 
17 & 2366 & 0.027 & 2450000.55 & 0.13 & 0.996 && 24 & 2166 & 0.234 & 2449999.78 & 0.65 & 0.978 \\ 
17 & 4107 & 0.018 & 2450000.12 & 0.42 & 0.985 && 24 & 2463 & 0.001 & 2450002.76 & 0.24 & 0.991 \\ 
18 &  667 & 0.751 & 2450006.28 & 0.52 & 1.000 && 25 &  425 & 0.031 & 2450003.26 & 0.32 & 1.000 \\ 
18 & 1201 & 0.123 & 2449999.32 & 0.34 & 0.983 && 25 & 2256 & 0.309 & 2450003.19 & 0.55 & 0.972 \\ 
18 & 2492 & 0.078 & 2450003.06 & 0.26 & 1.000 && 25 & 2800 & 0.069 & 2450001.64 & 0.36 & 1.000 \\ 
18 & 3232 & 0.064 & 2450000.77 & 0.15 & 0.980 && 25 & 2836 & 0.084 & 2450001.06 & 0.24 & 0.962 \\ 
18 & 3834 & 0.011 & 2450001.89 & 0.20 & 0.998 && 26 &  572 & 0.089 & 2450004.20 & 0.92 & 0.995 \\ 
18 & 3942 & 0.077 & 2450004.24 & 0.35 & 0.989 && 26 &  946 & 0.040 & 2450000.45 & 0.19 & 0.990 \\ 
18 & 4093 & 0.098 & 2449998.71 & 0.36 & 0.994 && 26 & 1376 & 0.001 & 2449999.56 & 0.22 & 0.997 \\ 
18 & 4246 & 0.115 & 2450006.97 & 0.62 & 0.980 && 26 & 3703 & 0.319 & 2450001.69 & 0.10 & 0.999 \\ 
18 & 4368 & 0.014 & 2450002.51 & 0.17 & 0.993 && 26 & 3875 & 0.003 & 2450001.59 & 0.09 & 1.000 \\ 
19 & 3463 & 0.075 & 2450000.00 & 0.21 & 0.995 && 26 & 4375 & 0.001 & 2450002.85 & 0.19 & 0.993 \\ 
19 & 4115 & 0.017 & 2450001.19 & 0.21 & 0.997 && 26 & 4395 & 0.002 & 2450001.02 & 0.20 & 0.999 \\ 
20 &  508 & 0.236 & 2450006.39 & 0.70 & 0.978 && 27 &  323 & 0.117 & 2450000.94 & 0.08 & 0.998 \\ 
20 &  625 & 0.038 & 2450002.39 & 0.18 & 0.981 && 27 &  646 & 0.012 & 2450000.57 & 0.38 & 0.968 \\ 
20 &  726 & 0.106 & 2450000.04 & 0.36 & 0.933 && 27 & 1512 & 0.034 & 2450000.35 & 0.37 & 0.999 \\ 
20 &  764 & 0.098 & 2450006.14 & 0.47 & 0.990 && 27 & 1587 & 0.578 & 2450001.80 & 0.20 & 0.992 \\ 
20 & 1218 & 0.009 & 2450000.64 & 0.41 & 0.994 && 27 & 2801 & 0.079 & 2450004.85 & 0.35 & 0.992 \\ 
20 & 3228 & 0.069 & 2450001.66 & 0.26 & 0.976 && 28 &  974 & 0.009 & 2450001.45 & 0.21 & 1.000 \\ 
20 & 3507 & 0.006 & 2450001.63 & 0.28 & 0.973 && 28 & 1355 & 0.005 & 2450001.30 & 0.18 & 0.998 \\ 
20 & 3703 & 0.027 & 2450003.29 & 0.45 & 1.000 && 29 &  491 & 0.396 & 2450002.76 & 0.38 & 0.931 \\ 
20 & 3831 & 0.032 & 2449999.51 & 0.12 & 0.998 && 29 &  943 & 0.085 & 2450008.30 & 0.49 & 0.999 \\ 
20 & 4260 & 0.065 & 2450002.41 & 0.28 & 1.000 && 29 & 1001 & 0.002 & 2450002.67 & 0.33 & 0.992 \\ 
20 & 5428 & 0.056 & 2450003.17 & 0.19 & 0.993 && 29 & 1063 & 0.013 & 2450001.85 & 0.28 & 0.979 \\ 
21 &  564 & 0.031 & 2450001.79 & 0.32 & 0.984 && 29 & 1149 & 0.071 & 2450001.20 & 0.60 & 0.980 \\ 
21 & 1563 & 0.041 & 2450022.50 & 1.68 & 1.000 && 29 & 1755 & 0.186 & 2450005.32 & 1.46 & 0.929 \\ 
21 & 1588 & 0.007 & 2450001.60 & 0.37 & 0.941 && 30 &  983 & 0.084 & 2450024.31 & 1.17 & 0.999 \\ 
21 & 2360 & 0.304 & 2450000.75 & 0.57 & 0.909 && 30 & 1034 & 0.240 & 2450000.80 & 0.34 & 0.996 \\ 
21 & 2601 & 0.066 & 2450000.03 & 0.23 & 0.987 && 30 & 1558 & 0.431 & 2450001.29 & 0.24 & 0.996 \\ 
21 & 2750 & 0.009 & 2450002.00 & 0.23 & 0.996 && 30 & 2126 & 0.215 & 2450000.72 & 0.32 & 0.993 \\ 
21 & 3596 & 0.034 & 2450003.44 & 0.49 & 1.000 && 30 & 2158 & 0.104 & 2449999.72 & 0.27 & 0.990 \\ 
\noalign{\vskip3pt}
\cline{1-6}  \cline{8-13}}
\setcounter{table}{1}
\MakeTableSepp{c@{\hspace{3pt}}r@{\hspace{7pt}}c@{\hspace{7pt}}c@{\hspace{7pt}}c@{\hspace{7pt}}c c@{\hspace{5pt}} c@{\hspace{3pt}}r@{\hspace{7pt}}c@{\hspace{7pt}}c@{\hspace{7pt}}c@{\hspace{7pt}}c}{12.5cm}{Concluded}
{\cline{1-6}  \cline{8-13}
\noalign{\vskip3pt}
Field& \multicolumn{1}{c}{ID} & $e$ & $E_2$ & $t_1$ [d] & $\sin i$ &&Field& \multicolumn{1}{c}{ID} & $e$ & $E_2$ & $t_1$ [d] & $\sin i$ \\
\noalign{\vskip3pt}
\cline{1-6}  \cline{8-13}
\noalign{\vskip3pt}
30 & 2508 & 0.080 & 2450002.46 & 0.54 & 0.993 && 37 & 2068 & 0.021 & 2449999.81 & 0.45 & 1.000 \\ 
30 & 3167 & 0.034 & 2449999.87 & 0.17 & 0.997 && 37 & 2768 & 0.290 & 2450000.55 & 0.71 & 0.963 \\  
30 & 5566 & 0.015 & 2450002.92 & 0.26 & 1.000 && 37 & 3186 & 0.113 & 2450003.36 & 0.50 & 0.989 \\  
30 & 5978 & 0.140 & 2450009.17 & 0.56 & 1.000 && 37 & 3240 & 0.033 & 2449999.55 & 0.25 & 0.998 \\  
31 &  399 & 0.297 & 2450002.57 & 0.31 & 0.989 && 37 & 4541 & 0.255 & 2450003.12 & 0.40 & 0.957 \\  
31 & 2277 & 0.039 & 2450001.96 & 0.31 & 0.999 && 37 & 6333 & 0.006 & 2450000.15 & 0.27 & 0.995 \\  
31 & 2670 & 0.034 & 2450000.14 & 0.30 & 0.998 && 37 & 6549 & 0.033 & 2449999.94 & 0.23 & 0.994 \\  
31 & 2793 & 0.065 & 2450001.42 & 0.41 & 0.996 && 37 & 7240 & 0.043 & 2450000.94 & 0.19 & 0.999 \\  
31 & 3308 & 0.020 & 2450000.84 & 0.31 & 0.983 && 37 & 8237 & 0.176 & 2450003.66 & 0.53 & 0.990 \\  
31 & 4092 & 0.207 & 2450002.18 & 0.29 & 0.970 && 38 & 1003 & 0.092 & 2450002.94 & 0.25 & 0.986 \\  
31 & 4328 & 0.025 & 2450011.57 & 0.26 & 1.000 && 38 & 1675 & 0.065 & 2450000.27 & 0.26 & 0.992 \\  
31 & 4787 & 0.276 & 2450002.92 & 0.41 & 0.986 && 38 & 3532 & 0.395 & 2450002.55 & 0.72 & 0.985 \\  
32 &  926 & 0.002 & 2450008.44 & 0.47 & 1.000 && 38 & 3758 & 0.011 & 2450000.43 & 0.34 & 0.990 \\  
32 & 1504 & 0.062 & 2450003.01 & 0.57 & 0.998 && 38 & 4718 & 0.102 & 2450002.50 & 0.48 & 0.982 \\  
32 & 1928 & 0.063 & 2450000.06 & 0.60 & 0.945 && 38 & 5059 & 0.081 & 2450002.76 & 0.12 & 0.997 \\  
32 & 2053 & 0.009 & 2450005.37 & 0.41 & 0.999 && 39 &  511 & 0.084 & 2450000.16 & 0.23 & 0.984 \\  
32 & 3061 & 0.016 & 2450000.14 & 0.45 & 0.996 && 39 & 1893 & 0.124 & 2450000.19 & 0.31 & 1.000 \\  
32 & 3753 & 0.045 & 2449995.62 & 0.58 & 1.000 && 39 & 2555 & 0.069 & 2449999.00 & 0.26 & 0.990 \\  
32 & 4589 & 0.068 & 2449997.85 & 0.27 & 0.999 && 39 & 2745 & 0.041 & 2450001.87 & 0.81 & 0.990 \\  
33 &  547 & 0.198 & 2450002.33 & 0.42 & 0.984 && 39 & 3008 & 0.043 & 2450001.65 & 0.31 & 0.995 \\  
33 &  686 & 0.001 & 2450002.17 & 0.21 & 1.000 && 39 & 4483 & 0.202 & 2450001.10 & 0.39 & 0.997 \\  
33 & 1629 & 0.009 & 2450000.96 & 0.20 & 0.992 && 39 & 5279 & 0.000 & 2450001.24 & 0.56 & 0.971 \\  
33 & 2278 & 0.119 & 2450001.35 & 0.25 & 0.981 && 39 & 5315 & 0.408 & 2450001.96 & 0.19 & 1.000 \\  
33 & 3188 & 0.140 & 2450003.61 & 0.52 & 0.998 && 39 & 6095 & 0.241 & 2450002.80 & 0.35 & 0.972 \\  
34 &  420 & 0.087 & 2450002.23 & 0.20 & 0.993 && 39 & 6554 & 0.040 & 2450001.05 & 0.21 & 1.000 \\  
34 &  793 & 0.143 & 2449997.29 & 0.78 & 0.963 && 40 & 1732 & 0.003 & 2450001.15 & 0.33 & 0.987 \\  
34 & 1615 & 0.059 & 2450002.31 & 0.30 & 0.988 && 40 & 1808 & 0.096 & 2450000.97 & 0.29 & 0.951 \\  
34 & 1706 & 0.045 & 2450000.92 & 0.38 & 0.990 && 41 &  207 & 0.319 & 2450006.21 & 1.29 & 0.974 \\  
34 & 1951 & 0.004 & 2450000.78 & 0.22 & 0.997 && 41 &  279 & 0.164 & 2450001.33 & 0.16 & 1.000 \\  
34 & 2456 & 0.092 & 2450003.99 & 0.67 & 0.995 && 41 & 1414 & 0.099 & 2450003.28 & 0.25 & 0.986 \\  
34 & 2580 & 0.030 & 2450002.19 & 0.26 & 1.000 && 41 & 1593 & 0.366 & 2450005.00 & 0.89 & 0.970 \\  
34 & 2823 & 0.127 & 2450001.38 & 0.18 & 0.999 && 42 &  872 & 0.043 & 2450002.10 & 0.16 & 0.997 \\  
34 & 2973 & 0.058 & 2450000.71 & 0.37 & 0.958 && 42 &  875 & 0.035 & 2450000.46 & 0.33 & 0.993 \\  
34 & 5744 & 0.018 & 2450002.74 & 0.27 & 1.000 && 42 & 1309 & 0.292 & 2449999.94 & 0.63 & 0.908 \\  
35 & 2440 & 0.214 & 2450001.24 & 0.20 & 0.999 && 42 & 4197 & 0.031 & 2449999.61 & 0.28 & 0.964 \\  
35 & 2508 & 0.066 & 2450015.41 & 1.44 & 0.999 && 42 & 4324 & 0.019 & 2450000.87 & 0.29 & 0.974 \\  
35 & 2897 & 0.010 & 2450000.72 & 0.26 & 0.996 && 43 &  305 & 0.073 & 2450005.41 & 0.55 & 0.994 \\  
35 & 4121 & 0.045 & 2450001.93 & 0.35 & 0.985 && 43 &  310 & 0.207 & 2450004.46 & 0.69 & 1.000 \\  
35 & 4556 & 0.005 & 2449998.35 & 0.26 & 0.995 && 43 &  834 & 0.087 & 2450001.81 & 0.25 & 0.992 \\  
35 & 4721 & 0.300 & 2450000.12 & 0.28 & 0.982 && 43 & 1023 & 0.189 & 2450000.19 & 0.28 & 0.977 \\  
36 & 1929 & 0.108 & 2450001.57 & 1.06 & 0.975 && 43 & 2852 & 0.010 & 2450002.65 & 0.40 & 0.998 \\  
36 & 3762 & 0.822 & 2450000.63 & 0.11 & 1.000 && 44 & 2264 & 0.014 & 2450001.75 & 0.16 & 0.999 \\  
36 & 4333 & 0.015 & 2450001.79 & 0.49 & 1.000 && 46 &  178 & 0.044 & 2449999.81 & 0.33 & 1.000 \\  
36 & 4493 & 0.010 & 2450000.32 & 0.29 & 0.996 && 46 & 1863 & 0.139 & 2450005.16 & 0.24 & 1.000 \\  
36 & 4494 & 0.007 & 2450000.33 & 0.31 & 0.998 && 46 & 1982 & 0.148 & 2450000.30 & 0.44 & 0.993 \\  
36 & 7720 & 0.183 & 2450003.00 & 0.24 & 1.000 && 46 & 2039 & 0.041 & 2450000.66 & 0.33 & 0.937 \\  
37 &  956 & 0.127 & 2450001.54 & 0.35 & 0.999 && 48 &  911 & 0.093 & 2450002.22 & 0.43 & 0.988 \\  
\noalign{\vskip3pt}
\cline{1-6}  \cline{8-13}}
\end{document}